\documentclass[%
reprint,
amsmath,amssymb,
aps,]{revtex4-1}
\usepackage{graphicx}
\usepackage{dcolumn}
\usepackage{bm}
\begin{document}
%\noindent
%Dear Editor,\hfil\hfil\break
%\noindent
%Dear Referee,\hfil\hfil\break
%\indent
%please find my reply to the referee's report at the end of this document.
%\hfil\hfil\break\noindent
%Sincerely, \hfil\hfil\break\noindent
%Gilles Couture\hfil\hfil\break\noindent

%\vfil\vfil\eject
%\vfil\vfil\eject

\title{Solar wind bremsstrahlung off DM in our Solar System \\
at 21 cm, 3 mm, and 12 microns Wavelengths}
\author{Gilles Couture}
\affiliation{D\'epartement des Sciences de la Terre et de l'Atmosph\`ere\\
Universit\'e du Qu\'ebec \`a Montr\'eal\\
201 Ave. du Pr\'esident Kennedy, Montr\'eal, QC\\
Canada H2X 3Y7}
\begin{abstract}
\noindent
Using Monte Carlo techniques, we calculate the bremsstrahlung spectrum 
that protons and electrons in the solar wind would produce when encountering 
dark matter (DM) particles within the solar system. We consider two 
types of interactions for the DM: one where a spin-1/2 neutral DM particle 
interacts with matter through the exchange of a lighter neutral scalar DM
particle and one where a scalar neutral DM particle interacts directly with 
the matter particle. We consider three wavelengths: 21 cm, 3 mm, and 12 microns. 
In order to estimate the significance of the signal, we compare our results to 
experimental results from ARCADE2 and PLANCK, and consider the sensitivities 
of three telescopes: VLA, ALMA and KECK. We find that in order 
to produce an observable signal in the first scenario, the neutral fermion must 
have a mass comparable to the solar wind particle while the 
exchanged particle must have a mass typically $few\times 10^5$ smaller. The scalar direct 
interaction however appears beyond reach for any reasonable coupling. 
\end{abstract}
\pacs{to come}

\maketitle

\noindent{\bf {Introduction}\hfil\hfil}
\hfil\hfil\break\noindent
Up to now, dark matter (DM) has remained one of the most elusive elements of modern physics. 
Although very well justified as an explanation for unaccountable observations about galaxy 
clusters for almost a century through the works of Zwicky [1] and, 
later, by Rubin and Ford [2] on the rotation of galaxies, 
later confirmed by numerous similar studies [3-4-5-6], 
it has failed to manifest itself in experimental searches. The exact nature of DM itself is 
still an open question as several variations [7] on the main theme of interaction with ordinary 
matter almost exclusively via the gravitational force, electrical neutrality, possible self 
interaction, and fairly large abundancy in the universe have been put forward over the years: 
sterile neutrinos[8], the lightest supersymmetric particle, such as the neutralino [9], 
WIMPS [10-11-12-13-14], exotic quark objects [15], axion [16-17-18], scalar/pseudo-scalar/vector 
[19] particles just to list a few, are all viable DM candidates.

Several avenues are used to try to determine the presence and properties of DM. 
Direct searches aim at detecting the interaction of DM with a liquid or solid detector, 
be it through a flash of light [20], nuclear recoil, [21-22-23] or a small deposit of energy 
[24-25]; these experiments require extremely high purity materials as any 
radioactivity could mimic the signal. 
As direct searches require an interaction between the DM particle and a nucleus, their sensitivity 
drops as the DM masse decreases and the interaction of these lower mass DM particles becomes more 
efficient with the electronic cloud [26-27]. 
In an indirect approach, one uses the fact that DM could accumulate around a massive object such 
as the Earth, the Sun, or the center of the galaxy. As they accumulate, the probability that they 
interact with each other or with matter particles increases, which could result into 
annihilation or scatterings whose signature would be an axcess of neutrinos, 
or cosmic or gamma rays produced 
in the Earth, the Sun, or the galactic halo [28-29-30] and detected on Earth. In collider 
experiments, one expects to produce DM particles directly and detect them indirectly 
through the observation of one or more jets accompanied by missing energy and/or 
missing $P_T$ as they will not interact with the detector. [31-32-33]
There is also the possibility that accumulation of DM particles could produce topological defects 
small on galactic scales but much larger than the Earth; such defects could produce signals (such 
as loss of synchronicity of atomic clocks) across the Earth as it passes through them. [34-35]

There has been some first observations over time, but none has passed convincingly  
the test of reproducibility up to now [36-37-38-39], [40-41-42].

Large scale numerical simulations and cosmological surveys [43-44-45-46-47] 
have confirmed the need for DM in order to reproduce the 
structures observed in the Universe based on accepted and tested gravitional laws.
Although DM is an interesting solution to the aforementioned issues, it is also possible 
to modify Newton's gravitational law on large, galactic scales while leaving it 
intact at smaller scales such as our solar system, in order to accommodate the 
rotational curves of spiral galaxies. [48-49-50]. This approach has received  
some experimental support recently as a study indicates [51] that the rotational curve 
of galaxies remains flat up to distances of order 1 Mpc, much larger than what was done before 
using the 21 cm line.

In this paper, we expand on a previous work [52] where we calculated the bremsstrahlung spectrum produced by protons in the solar wind when they collide with DM particles within the solar system. We considered a process where the DM particle is a neutral fermion that exchanged a lighter neutral scalar with the solar fermion. 
We concentrated on the visible and extrapolated slightly below. As we are dealing with a bremsstrahlung process, we expect a $E_\gamma^{-1}$ spectrum, which could lead to interesting rates at very low energy regimes. We will then consider three wavelengths: the 21 cm line 
(which corresponds to $6\times 10^{-6}~eV$) since it is very important in astronomy 
and is extensively studied [53], a second range in the $mm$ and finally a third 
one in the micron range. A first simple extension will be to consider not only 
protons but also electrons in the process. This opens up another mass range for the 
DM particles as the bremsstrahlung process is more efficient, as we will see, when 
all masses are comparable; we will not take into account the less numerous light ions 
of the solar wind although they would be relatively well described by the proton. We 
will also consider another type of process where the solar fermions interact directly with the scalar DM particle in a Higgs-like process.

For completeness, we will discuss briefly some technical issues before presenting 
our results; more complete discussions can be found in [52] We will explain briefly how 
to modify a Monte Carlo (MC) routine in order to concentrate on very low energy photons. 
In order to estimate the bounds one can put on the processes at hand, we will compare the fluxes produced by this bremsstrahlung process to current experimental sensitivities from ARCADE2 and PLANCK 
Collaborations as well as to the sensitivities of three telescopes: VLA (21 cm), ALMA (3 mm), and 
KECK (11.7 $\mu m$). 

\hfil\hfil\break
\noindent{\bf {Process}\hfil\hfil}
\break\noindent
We consider a simple bremsstrahlung process [54-55] where a charged particle 
(proton or electron in the solar wind) scatters off a neutral particle (dark matter,DM).

The masses and coupling of the DM particles are free. 
The process considered is then:
\begin{equation}
p(p) + q(q) \to p(\bar p) + q(\bar q) + \gamma(k)
\end{equation}
\noindent
where $p$ is our solar wind and $q$ is our DM. We do not take into account the
spin: we consider only a simple charged particle that emits a photon
when it scatters off a neutral particle.

In general, the presence of the Sun could increase the DM density in its immediate neighbourhood; 
how much of an enhancement above the galactic density is still debated [56-57-58-59-60] 
and difficult to settle as DM has not been observed yet.
DM particles could also accumulate inside the Sun and this accumulation could have two 
consequences: firstly, this could affect the sun itself and modify, for example 
its neutrino spectrum [61-62-63] and, secondly, DM could annihilate inside the sun and produce 
relativistic byproducts such as neutrinos that could be detected on Earth [64].  
We will simply take the DM density constant at $0.3~GeV/cc$ [65-66-67] 
throughout the solar system.

%As before, 
We set the speed
of the solar wind as 600 km/sec ($\beta_{_M} = 0.002$), and assume that the
DM is at rest in the galaxy; its motion being due to the motion
of the sun within the galaxy, which we take as 300 km/sec ($\beta_{_{DM}} = 0.001$).  
In the non relativistic limit, the maximum energy of the photon is
\begin{equation}
E_\gamma^{max} =
{(1/2)(M_{_M}~M_{_{DM}})(\vec\beta_{_M} -\vec\beta_{_{DM}})^2\over M_{_M}+M_{_{DM}}}
\end{equation}
\noindent
where $\beta_{_M}$ and $\beta_{_{DM}}$ are the usual relativistic parameters. 
This energy can reach several keV if we assume $M_{DM}\sim M_M = 1GeV$ and a few eV 
if we assume $M_{DM}\sim M_M = 511 keV$. 
The spectrum that we observe is similar to that of the Bethe-Heitler process [54]: 
$\sigma\sim E_\gamma^{-1}$ alost up to the maximum photon energy.   
%when plotting $d\sigma/dE_\gamma~vs~E_\gamma$ on a log-log scale, we obtain 
%a straight line over a very wide section of the spectrum, almost up to
%the maximum photon energy. 
As the divergencies encountered in bremsstrahlung 
processes when the photon energy goes to zero are cancelled by the higher order corrections [68-69-70] we should not have any problem since our 
photon energies, although very small, are finite.

\noindent{\bf {Monte Carlo (MC)}\hfil\hfil}
\hfil\hfil\break\noindent
In colliders, the collisions are always head-on as these maximize the 
center-of-mass energy (CME) and luminosities for given 
particle velocities and beam densities; the final boost to the laboratory
frame is simply along the z-axis, the x and y axes being the transverse axes.
In our process, as the solar wind will not necessarily collide head-on with the 
DM particle we modified the MC routine [52] to allow for collisions at 
any angle. In these conditions, the CME becomes 
\begin{equation}
s = M_{_M}^2+M_{_{DM}}^2 + 
2 M_{_{M}}M_{_{DM}}\gamma_{_{M}}\gamma_{_{DM}}(1+\beta_{_{M}}\beta_{_{DM}} 
cos(\theta_{_{q}}))
\end{equation}
\noindent
where $\theta_{_{q}}$ is the colliding angle and is defined such that 
$\theta_{_{q}}=0$ corresponds to a head-on collision (HOC) and the solar wind 
particle propagates along the positive $z-axis$.
Finally, one needs an extra boost matrix to the laboratory frame. 
We have verified that our routine conserves energy and total momentum at every step.

In a collision process, the heart of a MC is to consider the 
CME as a particle at rest whose mass is the CME and 
then let it decay into a first random mass particle and the first out-going particle; 
the energies and momenta of this decay, being a two-body decay, are all well 
defined in terms of the three masses involved. The MC then goes to the rest frame of 
the first random mass particle and lets it decay into a second random mass particle 
and the second outgoing particle; and so on until all the outgoing particles are 
produced. Since we have only three out going particles, we will have only one 
random mass particle that will decay into the last two out going particles. 
Of course, all these four-momenta have to be boosted back to the laboratory 
frame where the experiment is done. [55-71-72]

As we are interested in photon energies that are 6-9 orders of magnitude smaller than the  maximum photon energy allowed by kinematics ({\it ie} $6\times 10^{-6}~eV$ {\it vs} several keV for the proton and several eV for the electron), if we scan 
the whole parameter space, (as we did in [52]) very, very, few 
photons produced in the MC will have the energies we are interested in (roughly, a 
few events for every billion events produced). It will be extremely difficult to build any distribution. In order to produce very low energy photons, we have to modify the MC routine: as the photon is the second particle produced in 
the chain, we allow only very low masses for the first random mass particle and 
instead of scanning uniformly from 0 to 1, we scan from 0 to RMAX. 
Once we have all the necessary out-going momenta, we can calculate the 
$\vert amplitude\vert^2$; we must multiply it by RMAX since we have 
integrated only on a small portion of the parameter space and a low energy photon 
can be produced by a random mass particle of any mass. We have verified that this procedure gives coherent results as we vary RMAX.

\noindent{\bf {Photon mass}}
\hfil\hfil\break\noindent
As explained in [52], in order to avoid numerical instabilities coming from the negative $\vert amplitude\vert^2$ that can arise 
because of the null photon mass that can be numerically negative,
we treat the photon as a spin-1  massive boson with a very small mass
and whose summation over polarization states is $g^{\mu\nu}-k^\mu k^\nu/m_\gamma^2$. We will pick 
the photon mass very small and, as we will see later, varying its 
value by an order of magnitude does not affect the results noticeably.

\noindent{\bf {Interactions}\hfil\hfil}
\hfil\hfil\break\noindent
The first interaction is between a matter fermion (proton or electron) and
a neutral spin-1/2 DM particle and proceeds through the exchange of a neutral
scalar particle. The bremsstrahlung leads to two Feynman diagrams
in a t-channel process. In the limit where we can neglect the energy of the
photon compared to the masses of the particles, the overall cross-section
goes like $\lambda^4/m_{exch}^4$ where $\lambda$ is the coupling between
the two fermions and the scalar and $m_{exch}$ is the mass of the exchanged
scalar particle.

The second process is the interaction between the matter fermion and
the scalar particle in a Higgs-like process. Without the emission of a photon,
this type of interaction proceeds via two Feynman diagrams: 
one s-type and one t-type. The exchange particle in the t-type and 
the center-of-mass particle in the s-type is the matter particle. 
As the bremsstrahlung photon can be attached to any charged leg of the
diagrams, we have 6 diagrams to consider; therefore 21 terms in the 
$\vert amplitude\vert^2$. Symmetries and cyclicity of the trace are such 
that these 21 terms are of only three forms when rearranging the masses 
and the momenta appropriately. In this process, as the charged particle 
follows through in an s- or a t-channel, one can encounter some resonances as the propagator 
in the t-channel can have the form 
($m_p$ is the mass of the matter particle and $m_q$ is the mass of the DM particle) 
$(q-\bar p)^2 - m_p^2 = m_q^2 - 2 m_pm_q$
where the second expression is in the limit where we neglect the velocities of the particles and the energy of the photon. This will produce a resonance
at $m_q = 2 m_p$. When we take into account the
velocities of the particles and the energy of the photon, the resonance will
be shifted slightly, but it should still remain. This is shown in Table 1, for the case where the charged particle is an electron. As one can see, the
resonance is very, very narrow. 
We decided to not take into account these resonances as they would correspond to a 
single point in the available parameter space and would not be representative of the 
process at hand. We used instead reasonable masses based on a sampling of the 
available phase space. 

\noindent{\bf {Parameter Space}\hfil\hfil}
\hfil\hfil\break\noindent
In Tables 2A, 2B, 2C, and 2D, we give the ratios of the cross-sections as we vary
some parameters when taking a given combination of masses 
(matter-DM-exchange) as a standard. We consider a head-on collision as it is 
representative of the behaviour of the cross-sections at other colliding angles. 
Tables 2A and 2B refer to the Higgs-like scenario, while Tables 2C and 2D refer 
to the scenario where we have an exchange particle besides the DM particle. 
We consider only one photon energy regime as these ratios should be nearly constant 
for all energy regimes. It is important to consider that the
photon flux that will reach Earth does not depend only on the production
cross-section but also on the luminosity of the {\it cosmic collider}. As we will 
discuss shortly, the luminosity depends linearly on the density of the DM particles 
in the cosmos. This is why in these tables, we give two numbers: the 
ratio of the production cross-sections, $R_\sigma = \sigma/\sigma_{standard}$
and the ratio $R_{\sigma\rho} = (\sigma\times\rho/(\sigma_{standard}\times\rho_{standard})$ where 
we take into account the variation of the density ($\rho$) of the DM particles with their mass. We see that even though
the bremsstrahlung cross-section diminishes as the mass of the DM particle becomes
smaller, (a heavy charged particle will not emit much when colliding with a very light
object) the increase in the density of the DM particle can compensate in parts and make the process interesting; there is clearly an optimal combination of masses. 
We also note that the cross-section increases as the velocity of the DM particle increases, as expected. 

\noindent{\bf {Smoothing}\hfil\hfil}
\hfil\hfil\break\noindent
As explained in [52], 
%there are several distributions of interest, like 
%$ d\sigma/dcos(\theta_\gamma),
%d\sigma/dE_\gamma,~{\rm %and~mostly}~d^2\sigma/dcos(\theta_\gamma)~dE_\gamma$; where $E_\gamma$ is 
%the energy of the emitted photon and $\theta_\gamma$ is the photon angle %with respect 
%to the incoming matter particle. 
the energy distribution converges relatively quickly, but the angular 
distribution, as a photon of any energy
can be emitted at any angle since we have a three-body process,
is slower to converge and required $2\times 10^9$ events to reach
a relatively clean distribution. Once we had these relatively clean distributions, 
we used a fifth-order smoothing algorithm [73]:
%{\it \`a la} Savistsky-Golay [72] to smooth them. The algorithm we used
%was fifth order:
\begin{equation}
y_i = {1\over 35}(-3 y_{i-2} + 12 y_{i-1} + 17 y_i + 12 y_{i+1} -3 y_{i+2})
\end{equation}
\noindent
and after 150 iterations, the spread was reduced by about a factor 5.

\noindent{\bf {Luminosity}\hfil\hfil}
\hfil\hfil\break\noindent
The beam of the cosmic collider we are considering is one meter by one meter in width and its length is the length that the colliding particles travel in one second: 300000 m for the DM particle and 600000 m for the solar wind particle. On astronomical scales, we consider that such a collision volume is in fact a point in space. 
In these conditions the luminosity [74-75] is given by 
\begin{equation}
{\cal L} = \rho_{_{M}}\rho_{_{DM}} L_x L_y v_{_{m}} v_{_{DM}}\Delta t~K
\end{equation}
\noindent
where $L_x$ and $L_y$ are the transverse dimensions of the beam (which we take as 1 meter), 
$\Delta t$ is the time scale of the collision (which we take at 1 second), and the $K$-factor  
takes into account the non-collinearity of the beams:
\begin{equation}
K = \sqrt{(\vec\beta_{_{M}}-\vec\beta_{_{DM}})^2 - 
 (\vec\beta_{_{M}}\times\vec\beta_{_{DM}})^2}
\end{equation}
\noindent
The DM density is constant throughout the solar system, but the matter density varies as we move away from the Sun ans is given by 
\begin{equation}
\rho_{_{M}}={{{\cal R}_{_{sw}}/m_{proton}}\over 4 \pi D^2 v_{_{m}}}
\end{equation}
\noindent
where ${{\cal R}_{_sm}}$ is the emission rate of the solar wind, which we take as 
$10^9$ kg/sec and $D$ is the distance between the colliding point and the Sun (which we take as a point). We assume that the solar wind is mostly made of protons and will assume that the electron density in the solar wind is that of the proton.

Within the conical observation volume that we consider, several such 
cosmic collisions can happen every second: the number of interacting colliders is given by the ratio (volume of the observation cone)/(volume of the cosmic collider) and the volume of the cosmic collider is $L_x\cdot L_y\cdot (600000 + 300000)$. We neglect the volume of the interacting region and the special case where the collision is backwards (as opposed to head-on) where the two beams would overlap.

\noindent{\bf {Sampling the Solar System}\hfil\hfil}
\hfil\hfil\break\noindent
As geometrical details were given in [52], we will simply say that 
the volume where the collisions between the solar wind and the DM particles take place is a very long cone that will extend to several au and whose opening angle represents the field of view of the observer. As the collision angle between the DM 
and the matter particles will vary with our position within the cone, we considered 
18 colliding angles: from 0 (HOC) to 180 degrees, by increment of 10 degrees. 
We verified that both $d\sigma/dE_\gamma$ and 
$d^2\sigma\over dcos(\theta_\gamma) dE_\gamma$ have the $E_\gamma^{-1}$ behaviour 
over several orders of magnitude since they are bremsstrahlung. 
As can be seen in Tables 2A-2D, varying the mass of the photon by an order
of magnitude has no numerical consequences on the results.

The important distribution has the form
\begin{equation}
{d^2\sigma\over dcos(\theta_\gamma) dE_\gamma} \sim {A(\theta_q) + 
B(\theta_q)cos(\theta_\gamma)\over E_\gamma} 
\end{equation}
\noindent
where the parameters $A$ and $B$ vary with $\theta_q$ as 
the total cross-section varies with $\theta_q$. 
On Figure 1 we plot a very thin slice across one such distribution: for a given value of $\theta_q$ 
and we set $\theta_\gamma = 0$.  Clearly this distribution has the $E_\gamma^{-1}$ signature 
over several orders of magnitude. Furthermore, this figure is in fact two curves: the first 
one is at low energy and runs from about $2\times 10^{-6}~eV$ to about $5\times 10^{-3}~eV$ and the 
second one is for higher energies and runs from about $1.5\times 10^{-4}~eV$ to about $0.3$~eV
The fact that these two lines overlap very nicely as they should gives us confidence that the MC procedure 
that we outlined above is sound. Note also that in order to have a manageable file, we plot only a 
small fraction of the total number of points produced.

In order to integrate over the volume of interaction we pick the point of interaction 
at random over the selected volume. Once this point is chosen, both $\theta_q$ and 
$\theta_\gamma$ are fixed since the photon has to go to Earth and we assume that 
the solar wind moves radially from the sun; at this point, we do not take into 
consideration the effects of the solar magnetic field. 
In scanning the volume around the Earth, we chose a cone whose apex is the Earth and
opening angle is 1 degree (total angle of 2 degrees and total solid angle of
0.001 sr) As the opening angle is small, we observe that the amplitude of the 
signal behaves as $\theta^2$ as we increase the opening angle. 
One can also vary the height of the cone, and as it is varied, we see a rise in the photon production rate and then the signal saturates once we reach a 
certain height. Considering that the density of the solar wind will decrease as we 
go farther and farther from the sun, this is expected. We also see a decrease in the 
signal for a given number of sampling events generated as it becomes more and more 
difficult to sample properly a larger and larger volume. 
We compromised on 10 au as a good height of the cone where the computing time was 
reasonable and close enough to the saturation value to be a good estimate. 
One must sample the volume of the cone as uniformly as possible. As we did 
in [52] we used a parallepiped ($x-y-z$) and tailored it to the 
dimensions of the cone in order to keep as many interacting points produced as possible. We consider five angles for the orientation of the observer/detector: 5 degrees is where 
our back is towards the sun (around midnight); 45 degrees is where our back is partly 
to the sun; 90 degrees is where the sun is on the horizon; 135 degrees is where we 
are partly facing the sun and, as we do not want to face the sun directly, 175 degrees is where we are almost facing the sun (noon). 
Considering that the volume element in 3D is $r^2dr~dcos(\theta)~d\phi =
{1\over 3} dr^3~dcos(\theta)~d\phi$ one could also scan uniformly
$cos(\theta),~\phi$, and $r^3$. As $r^3$ can span a few orders of magnitude, one
has to be very careful in scanning it uniformly.

We will pick interaction points within the volume of the parallepiped as randomly 
and uniformly as possible and consider only those events that are within our cone 
of interest. Once the interaction point is picked, we calculate the luminosity 
at this point based on the DM density in the cosmos, the density of the 
solar wind as it moves radially from the sun, and their relative momenta. 
For each interaction point, we have to decide what will be the energy of the 
outgoing photon; this is through the previous distribution. Once the outgoing 
angle is known, there is an energy range for the photon energy: 
$E_\gamma^{min} < E_\gamma < E_\gamma^{max}$ 
We will scan this energy range randomly and associate the given cross-section 
to the event. Once we multiply it by the luminosity, we will have the number of 
photons associated with this particular event, it's configuration, and it's photon 
energy.

One has to be very careful when sampling the doubly differentiated 
distribution since it spans several orders of magnitude in photon energy ($E_\gamma$) 
and represents weighted events ({\it ie} a cross-section is associated to this event)

We used a two-step process. A first random number will tell us which order of magnitude 
we will sample and a second random number will tell us which unit within this order 
of magnitude will be sampled. As all our bins have the same width, we have 10 times more 
bins as we go from one order of magnitude to the next. To properly account for the 
number of events distributed within each order of magnitude, we have to have the 
same number of bins in each order of magnitude of photon energy. Therefore, we must 
add up some bins in a given order so that each bin has the same number of events 
as the bins of the previous order, within normal variance due to MC technique. 
We loose some precision in the higher energies, but this procedure preserves  
the main features of the distributions.

Another possible procedure would be to have a given number of bins equally distributed among the different energy scales, assign a number to each one of 
these bins and sample them randomly. For example, 100 bins between $10^{-7}$ and 
$10^{-6}$ eV, 100 bins between $10^{-6}$ and $10^{-5}$ eV and 100 bins between 
$10^{-5}$ and $10^{-4}$ eV. Then sample randomly between 1 and 300.

\noindent{\bf {Results}\hfil\hfil}
\hfil\hfil\break\noindent
On Figure 2, we present the result of this procedure for a given process where a solar wind electron 
encounters a DM particle whose mass is 1000 keV and they exchange a 100 keV scalar particle. This figure is composed of 6 independent lines. The bottom pair 
corresponds to an observation angle of 5 degrees,  
%({\it ie} with our back to the sun, around midnight) 
the middle two correspond to an observation angle of 135 degrees 
%({\it ie} at 45 degrees with respect to the sun) 
and the upper two correspond to an observation at 175 degrees. 
%({\it ie} almost facing the sun). 
The three lower energy curves span an energy range of $10^{-9}~keV$ to $10^{-7}~keV$ and the three upper energy curves cover the 
energy regime $10^{-7}~keV$ to $5\times 10^{-5}~keV$. They were produced by 6 different MC runs. As they represent the same process, one expects the lower 
energy curves to join continuously with their higher energy counterparts. 
The fact that they join rather nicely indicates that the method outlined above is sound. Unfortunately, we did not use this procedure in [52] which resulted in rates and slopes being overestimated; an {\it erratum} is in progress.

We note that the slopes of these curves are close to $-1$, varying from -0.95 to -0.97. This means that integrating over a certain volume of space does not change 
the  $1/E_\gamma$ behavior. Recall also that these curves correspond to an opening 
angle of 0.001 sr for the observation cone, which corresponds to a field of view of 
2 degrees in the sky.

All the processes that we studied produced figures similar to Figure 2. 
We will then simply use these distributions to get the event rate at a given photon energy; 
namely $6\times 10^{-9}~keV$ or $\lambda = 21 cm$
This is given in Table 3A and Table 3B where we present the number of photons one could 
expect to receive on Earth; the units are ${number~of~photons / (keV~m^2~sec)}$. Recall that 
in Table 3A, the process goes essentially like $\lambda/m_{exch}^4$ and while the mass of the 
exchange particle is much smaller in the electron scenario, so is the energies since 
we assume similar velocities for the electrons and for the protons. 
These smaller energies explain why the observation rates are 
comparable for the electron and the proton; the density is also important as it will be 
larger for the lighter DM particles. In Table 3B, the large difference in observation rates 
between the electrons and the protons come from the large propagators involved in the proton 
case ($\sim m_p^{-4}$) as opposed to the electron case ($\sim m_e^{-4}$). Here also the 
density of the DM particle plays a role.  

In order to transform the observation rates of Tables 3A and 3B into energy fluxes, we have 
to perform the following operations:
$
52\times 10^{-6}\times 0.5\times 10^{-9} \times 6\times 10^{-6}\times 1.602
\times 10^{-19}J/eV\times NB~/0.001 
$
which give us an energy flux of 
$
2.5\times 10^{-35}\times NB~Watt~m^{-2}sr^{-1}
$
where $52\times 10^{-6} keV^{-1}~m^{-2}~sec^{-1}$ is the number obtained from the upper-left 
line on figure 3 (175 degrees) at $6\times 10^{-9}~keV$,  
$0.5\times 10^{-9}~keV$ is the width of the bins at $6\times 10^{-9}~keV$, 
$6\times 10^{-6}$eV is the energy of a photon since we work at 21 cm, 
0.001 Sr is our solid angle, and $NB$ is the number of bins required to 
match the bandwidth of the detector on Earth. 
As the number of photons decreases rather slowly from one bin to the next, if the number of adjacent bins required is small, we will assume that 
the flux is constant over the bins.

Dividing these values by the width of the bin in $Hz$ will give us the specific intensity: 
$0.5\times 10^{-6}~eV\to 121~MHz$ which gives us $2\times 10^{-43}~W~m^{-2}~Hz^{-1}~Sr^{-1}$. 
Recall that a common unit is the $Jansky = 10^{-26}~W~m^{-2}~Hz^{-1}~(Jy)$.  Due to the $1/E_\gamma$ of the bremsstrahlung process, this number will 
be constant over the whole spectrum.

 As for the proton, we get a specific intensity of $3\times 10^{-45}~W~m^{-2}~Hz^{-1}~Sr^{-1}$ since the rate is $0.77\times 10^{-6}~keV^{-1}~m^{-2}~sec^{-1}$ and the bin is $1.45\times 10^{-9}~keV$ in width.

\noindent{\bf {Detectors, Sensitivities and Limits}\hfil\hfil}
\hfil\hfil\break\noindent 
Since the first process goes like $(\lambda/m_{exch})^4$ it offers some flexibility in the masses 
probed since the mass of the exchanged particle is free. Furthermore, when considering the protons and 
the electrons in the solar wind, this process probes very different DM mass domains since it is more 
efficient when the masses that collide are similar. The other 
process on the other hand offers no such flexibility since the particle that appears in the $s-$ 
and $t-$ channels is the matter fermion. Therefore, except for the very narrow resonances mentioned 
previously, the second process goes like $\lambda^4$.  
In what follows, we will consider only the DM+exchange scenario and defer discussion of the 
Higgs-like scenario to the section Sensitivity at 21 cm.

Unlike the optical sky where it is generally relatively easy to see a star against a dark sky [76], 
the radio sky has enormous backgrounds. These backgrounds that can mix with signals and cover them 
have been observed, studied, and modeled [77, 78, 79, 80, 81, 82, and 83, 84]
Some of the major backgrounds are synchrotron radiation produced by electrons (or other charged 
particles) of (relatively) high energies spinning in cosmic magnetic fields, from our galaxy or 
beyond [85,86,87] free-free emission from the encounter of charged particles scattering off 
other particles in the cosmos, [88,89] dust particles, spinning or not, heated by radiation [90] or the 
isotropic background radiation [91]. 
One of the objectives, of course is to produce precise and accurate 
maps of the sky and our galaxy in order to subtract them from the observed 
skies so that a faint signal can be extracted [92,93]

We will not perform such a detailed analysis here. Our approach will be instead to estimate the 
strongest and weakest limits we could put on the bremsstrahlung process studied here by using 
what could be described as a best case scenario and a more conservative or realistic one. 
We will use data from observations and from theoretical sensitivities from detectors on the ground.

\hfil\hfil\break\noindent{\bf Sensitivities through ARCADE2\hfil\hfil}
\break\noindent
As a first step in estimating the sensitivities of current data on this process, we 
will use the results of the ARCADE2 experiment. This Collaboration has measured an anomaly in the 
temperature of the sky [94] This anomaly has spurred a lot of interest and has been used, for example 
to constraint models of extra galactic emissions [95] or more exotic processes such as axion-photon 
conversion [96], bremsstrahlung emission in the early universe [97], axion spontaneous decay [92], 
or WIMPs [98]

Their best fit to their data is given as [94]
$$
T = 2.725\pm 0.001 + 24.1\pm 2.1~\big({\nu\over \nu_0}\big)^{-2.599\pm 0.026}
$$
\noindent
with $\nu_0 = 310MHz$.  The first term is the CMB temperature and the 
second term is an excess temperature that they observe from 22 MHz to 10 GHz. 
We will compare our signal to the power generated by this excess temperature. 
(Even though [99] quotes a slightly more precise value for the CMB temperature a 
$2.72548\pm 0.00057~K$, we will use the previous fit by the ARCADE2 Collaboration)
In order to translate this sky-temperature to antenna temperature and power observed, following ARCADE2, we use 
$$
T_A = {x~T\over e^x -1};~~x = {h\nu\over kT};~~P = {k~T_A~\nu^2\over c^2}
$$
\noindent
where
$h$ is Planck's constant and $k$ is Boltzmann's constant.

At $1.43~GHz$, we calculate $T = 3.178K$ and $T_A = 2.6908K$ 
which leads to a power of $98.584kJy~Sr^{-1}$. Assuming that the uncertainties are uncorrelated, 
we can combine them to get the maximum and 
minimum powers; we get $ P = 98.584\pm 2.1 kJy~Sr^{-1}$ Doing the same with 
a temperature of $2.725\pm 0.001~K$, we get $84.37\pm 0.03~kJySr^{-1}$ which 
brings us to an excess power of $12.2\to 16.4~ kJySr^{-1}$. This is to be  
compared to our bremsstrahlung power of 
$2\times 10^{-17}JySr^{-1}\times \big(\lambda/(m/100keV)\big)^4$; 
using the central value for the excess power leads 
to $\big(\lambda/(m/100keV)\big) > 1.5\times 10^5$. We see then, that 
the mass of the exchanged particle has to be in the eV range if we require that 
$\lambda\sim 1$

Combining the bandwidth of 18 MHz used by Reich and Reich [87] at 1.43GHz  
with the ARCADE2 temperature at that frequency, we get a power of $98.584\times 10^{-23}~Wm^{-2}Hz^{-1}Sr^{-1}\times 18MHz = 1.77\times 10^{-14}~Wm^{-2}sr^{-1}$ 
which is to be compared to our value of 
$2.5\times 10^{-35}~Wm^{-2}sr^{-1}\big(\lambda/(m/100keV)\big)^4$. This leads to 
$\big(\lambda/(m/100keV)\big)>1.6\times 10^5$ which is consistent with our previous bound.

Combining the uncertainties in the same way but in the 3.09-3.30 GHz bandwidth we obtain 
$P_0 = 414.63\pm 0.16 kJySr^{-1}$ and 
$P = 423.4\pm 1.7 kJySr^{-1}$ which leads to $6.9~kJysr^{-1} < P_{excess} < 10.7~kJySr^{-1}$ 
and $\big(\lambda/(m/100keV)\big ) > 1.46\times 10^5$ while the bandwidth $3.30-3.52 GHz$ 
gives $P_0 = 471.41\pm0.18~kJySr^{-1}$ and $P = 479.9\pm 1.7~kJySr^{-1}$ which leads to 
$6.7 kJySr^{-1} < P_{excess} < 10.4 kJySr^{-1}$ and 
$\big(\lambda/(m/100keV)\big) > 1.37 \times 10^5$

Using the bandwidth of 210 MHz, we calculate a power of $1.89\times 10^{-14}~Wm^{-2}Sr^{-1}$ 
which is to be compared to a power of $2.5\times 10^{-35}~Wm^{-2}Sr^{-1}$ 
from our figure 2. This leads to $\big(\lambda/(m/100keV)\big)>1.7\times 10^5$  and similarly for the 220 MHz bandwidth at $3.30-3.52~GHz$. 
 
At 10GHz, the same procedure gives $P = 3825.4\pm 4.0~kJysr^{-1}$ and $P_0 = 3821.1\pm 1.5~kJySr^{-1}$. As can be seen, in this scenario $P_{min}< P_0^{max}$; one should then 
rely instead on the central values which give $P_{excess} = 4.3~kJySr^{-1}$ and 
$\big(\lambda/(m/100keV)\big) > 1.2\times 10^5$.

Similarly at 50GHz and 100GHz, 
%(figure 5 of ARCADE2) 
$P_{min}<P_0^{max}$ and, relying  on 
the central values only, we obtain  
$\big(\lambda/(m/100keV)\big) > 1\times 10^5~{\rm and}~0.8\times 10^5$ 
One can see a trend: as the frequency increases, the constraints on $\lambda/m/100keV$ get 
slightly stronger. This is due to the $E_\gamma^{-1}$ behavior of the bremsstrahlung process: the energy does not decrease as fast as the current spectrum as the frequency increases. This also means that a constraint that would only saturate the signal at low 
frequencies would definitely produce a very strong signal at higher frequencies. 
%In other words, if we require that our signal be barely visible at low frequencies, it will be 
%stronger at higher frequencies. 
\hfil\hfil\break\noindent
{\bf EDGES\hfil\hfil\break}
\noindent
The EDGES Collaboration [100] has observed an anomalous dip at around 78 MHz: when they model the 
21 cm signal and residuals, there remains a dip of about 0.5K centered at 
about 78 MHz with a width of about $\pm 10~MHz$. This anomaly has also attracted a lot 
of attention.[101,102,103] One must mention however that the SARAS Collaboration [104] 
did not see such a dip; this dip still remains an open question. 
Their frequency range (50-100 MHz) corresponds to an 
energy range of $0.207\times 10^{-6}~eV - 0.414\times 10^{-6}eV$ and requires an extrapolation 
by a factor of 4 from our low energy results; still trustworthy. 
It would be difficult to explain this dip by a bremsstrahlung process 
such as the one studied here but let's assume that their 
fit is missing an element and we identify this element with the 
bremsstrahlung process considered here.  As their overall 
uncertainty is about 0.2K, this would translate into a power-gap of 
$1.87\times 10^{-23}~W~m^{-2}~Hz^{-1}~Sr^{-1}$ which then leads to 
$\big(\lambda/(m_{exch}/100keV)\big)> 0.3\times 10^5$. 

Again, it would be difficult to explain the dip observed by EDGES by a bremsstrahlung process but
their uncertainties give a good idea of the limits one could set on our parameters.

\noindent{\bf Sensitivities through PLANCK\hfil\hfil}
\break\noindent
The PLANCK Collaboration has published very extensive sky maps in the micro-wave regime. We will consider [105, 106] where 
they take into account several emission processes: synchrotron, free-free, spinning dust, thermal dust and thermal Sunyaev-Zeldovich. When they model their observations and combine them with 
other surveys, their fits agree with the data extremely well and they are left with residuals 
that are just a few $\mu K$ over most of the sky and about 1\% over the 
remaining small portions. A summary of their model and the measure of the 
different elements 
is given in their Figure 4 and Table 3. In their model, they neglected some sub-dominant processes such as magnetic dipole emission from dust grains. Furthermore, attempts at incorporating extra-Galactic point sources in the model by subtracting catalogued sources 
in pre-processing maps were unsuccessful in that no statistical gain was obtained in this manner. 
 A conservative approach 
is to consider that the bremsstrahlung signal that we calculated would have modified their 
fitting parameters if it had been strong enough. Therefore, considering that their residuals are 
about $4\mu K$, we set our signal at $20\mu K$. At 21 cm, this leads to an excess power of 
$6.3\times 10^{-27}Wm^{-2}Hz^{-1}Sr^{-1}$ and $(\lambda/m/100keV)>0.13\times 10^5$.  
A more conservative approach is to consider only the two dominant processes (synchrotron and 
free-free) that produce a temperature of about 1.5K together and ask that our signal be 10\% 
of this so that the fitting parameters would have been off. In this way, we get, at 21 cm an 
excess power of $4.7\times 10^{-23}Wm^{-2}Hz^{-1}Sr^{-1}$ and 
$\big(\lambda/m/100keV\big)>1.2\times 10^5$.

At 100GHz however, thermal dust becomes the 
dominant background with a temperature of about $33\mu K$ while synchrotron radiation accounts 
for about $4\mu K$ and free-free for at most $1-2\mu K$ in most scenarios. Therefore, we set our 
signal at about $150\mu K$, which leads to a power of $2.3\times 10^{-22}Wm^{-2}Hz^{-1}Sr^{-1}$ and $\big(\lambda/m/100keV\big)>1.8\times 10^5$.

\noindent 
{\bf Sensitivity at 21 cm}\hfil\hfil\break\noindent
Longair [77] states a sky intensity at 1.42 GHz of $2\times 10^{-21}Wm^{-2}Hz^{-1}Sr^{-1}$ which 
leads to a first limit of $\lambda/(m_{exch}/100keV)> 3\times 10^5$ if we assume that all the signal 
is due to the bremsstrahlung process. This could be considered the worst case scenario and a 
conservative limit.

The  VLA telescope [107,108] has a full width at half power of about 30 minutes of arc at 21 cm;
we will consider a field of view of 1 degree, which will cover most of the energy
captured by the detector. This would give us a solid angle of 0.24 mSr, compared to our 
1 mSr. The sensitivity is about $60~\mu~Jansky$ at a bandwidth of 600 MHz at 1420 MHz with 
an integration time of 1 minute. We calculate a sensitivity of 
$1.4\times 10^{-18} W~m^{-2}~Sr^{-1}$. From Table 3A, we see that, in the process DM+exchange we 
would require 
$1.21\times 10^{-34}\lambda^4/(m_{exchange}/100keV)^4 > 1.4\times 10^{-18}$ or 
$\lambda/(m_{exchange}/100keV) > 1.04\times 10^4$. 
Therefore, in order to begin to set limits on 
$\lambda\sim 1$, the mass of the exchange particle must be less than a few tens of eV while 
keeping the mass of the DM particle in the few hundreds of keV range. 
The same is true for the proton where the constraint becomes 
$\lambda/(m_{exch}/100MeV)> 4.4\times 10^4$ at 175 degrees: the mass of the exchange particle mut 
be less than a few tens of keV while keeping the mass of the DM particle in the several hundreds of 
MeV range. One notes that in this scenario, the constraints do not depend much on the observation 
angle due to the fourth power involved. Since the sensitivity of the detector is proportional to 
$1/\sqrt{t}$ one could improve this bound by integrating over a longer period of time (we used 1 
minute  here) but the gain would not be substantial, owing to the fourth power of the coupling.

Clearly, this is the best one could ever do since we assume that there is no background and the 
VLA detector measures down to its limit. There are also some detector issues that we neglected 
here such as side lobes.

\hfil\hfil\break\noindent
A more conservative approach at VLA is to consider that there is more or less constant background at 
21 cm of about $20 mJy/beam$ [78,83] away from the Milky Way, where $beam$ is the standard 
Gaussian definition $beam = \pi\theta_{min}\theta_{max}/4 ln(2)$ 
If we assume a simplified $beam$ that is 
a step function whose width is 1 degree of arc, we get a $beam$ of 0.00024 Sr, which leads to 
$8.3\times 10^{-25}Wm^{-2}Hz^{-1}Sr^{-1}$ for the average noise in the 21 cm band. 
If we require that our bremsstrahlung signal be twice as large as this noise, 
we then get $\lambda/(m_{exch}/100keV) > 0.5\times 10^5$ Since this represents a more or less constant 
background one could be more stringent and require that we compare two bremsstrahlung signals: one 
at 135 degrees and one at 90 degrees, for example, and subtract them in order to reduce the real 
background and be left with the bremsstrahlung signal. One would then loose about a factor of 10 in 
signal, which would lead to $\lambda/(m_{exch}/100keV)>\sim 1\times 10^5$. This would be a more 
realistic limit.

\noindent 
Higgs-like scenario\hfil\hfil\break\noindent
Considering Table 3B, when an electron interacts with a DM scalar particle in a Higgs-like 
scenario, we follow the same procedure as above, but the bins are a little bit smaller at 
about $0.2\times 10^{-6}eV$, which leads to $6.5\times 10^{-36}\lambda^4 > 1.4\times 10^{-18}$ 
at an observation angle of 175 degrees, or $\lambda > 2.15\times 10^4$. 
Similarly, $\lambda > 3.8\times 10^4, 4.6\times 10^4, 5.6\times 10^4, 5.7\times 10^4$ for 
$\theta = 135, 90, 45, 5~degrees$ respectively. In this scenario, as can be seen from 
Table 2A, varying the mass of the DM particle will not change our results by much since 
we are close to the most efficient combination. Since the sensitivity depends on 
$\sqrt{time}$, one would not gain very much by integrating over a longer period of time. 
Again, from Table 3B, we see that the rates are very much smaller when we consider the protons in 
the solar wind. This comes from the $m_{exch}^{-4}$ behaviour of the cross-section since the 
exchange particle in this scenario is the fermion itself, the proton. These rates are extremely 
small and such large couplings are clearly not physical; we must conclude then that this process is 
out of reach.

\noindent{\bf Sensitivity at 3 mm}\hfil\hfil\break\noindent
Considering the specifications of the ALMA detectors [109,110]: bandwidth from 84 to 117 GHz, and 
noise temperature of 60K, leads to $\Delta T_{RMS} = T_{sys}/\sqrt{\Delta\nu~\tau}$ and 
assuming that our signal is twice as large as the noise and the integration time is 60 sec 
$T_{sys}\sim 60K + 120 K$ leads to 
$\Delta T_{RMS} = 1.3\times 10^{-4}K$ and 
$kT_{RMS}/\lambda^2 = 20kJy = 2\times 10^{-22}$ and $\lambda/(m_{exch}/100keV)>1.6\times 10^5$.
This is the best case scenario where we consider only the 
ultimate sensitivity of the detector.

We can also proceed through the current Alma-Science-Primer [111] which states a sensitivity of 
0.082 mJy/beam in band 3 which ranges from 84-116 GHz, or 3.6-2.6 mm with a field of view of 
60'' - 50'' and takes an integration time of 60 seconds. Assuming a symmetrical FOV of 1 arc-min 
and the standard Gaussian beam where $beam = \pi\theta_{min}\theta_{max}/4 ln(2)$, we get a 
sensitivity of $3.4\times 10^{-23}Wm^{-2}Hz^{-1}Sr^{-1}$ which then leads to 
$\lambda/(m_{exch}/100keV)> 1.1\times 10^5$; which is quite consistent with the previous result. 
Again, this can be seen as the ideal scenario where all backgrounds are very 
well understood and accounted for.

In order to estimate the best one can do in filtering data, we will consider M82 whose diameter 
is about 40000LY and is about 13MLY away from Earth. Although the galaxy is rather elongated, 
assuming it is circular will be sufficient for our purposes. This then leads to a solid angle of 
about $7.4\times 10^{-6}Sr.$ If we consider the total amount of energy coming from M82, 
(A, M, B-2) at 3 mm (100GHz) 
it is about 1 Jy. This leads to an energy flux of 
$1.35\times 10^{-21}Wm^{-2}Hz^{-1}Sr^{-1}$ and $\lambda/(m_{exch}/100keV) > 9\times 10^5$. Clearly, M82 
represents a strong signal and requiring that the bremsstrahlung process be as strong as M82 
is rather conservative. We can conclude however that the bremsstrahlung process could lead to a limit 
of $\lambda/(m_{exch}/100keV)> 3-5\times 10^5$ in a realistic situation. 

%(NB 1 arc-min as I calculate it before leads to multiplying by 15000000 while using the expressions 
%$\Omega = \pi\Theta_{min}\Theta_{max}/(4ln(2))$ leads to multiplying by 42000000

Another measurement that gives a very good idea of the reachable limit is that of [81] where it is 
stated that the temperature of the sky is about $10~\mu K$ at 100 GHz. This leads to 
$\lambda/(m_{exch}/100keV) > 1\times 10^5$. We will consider this as the best that could be achieved 
in that regime.

Finally, a very conservative estimate is from Longair [77] which gave an overall signal 
of about $3\times 10^{-18} W~m^{-2}Hz^{-1}Sr^{-1}$ and leads to 
$\lambda/(m_{exch}/100keV)> 62\times 10^5$. This window has a very strong background and assuming 
that it all comes from the bremsstrahlung process is the most conservative one could do and not very realistic as a lot of it can be explained as synchrotron or free-free radiation.
\hfil\hfil\break\noindent

\noindent
{\bf Sensitivity at 11.7$\mu m$}\hfil\hfil\break\noindent
Our results allow us to extrapolate safely by a factor of 2 or so in order to reach the KECK 
detector, [112] which has a fairly large operational bandwidth. We will consider the 11.7 microns or 
$1.06\times 10^{-4}keV$ window. In the LWS Instrument Specifications, one sees that a sensitivity 
of 17 mJy gives a SNR = 1 with an integration time of 1 sec, centered at 11.7 microns 
and a bandwidth of 2.4 microns; this translates into a bandwidth of $10.5 - 12.9$ microns or 
$0.96\times 10^{-4}~keV<E_\gamma<1.18\times 10^{-4}~keV$ and a bin width of $2.2\times 10^{-5}~keV$ . 
This translates into about 
$1.25\times 10^6~photons~m^{-2}sec^{-1}$ Since the opening angle is $2.7\times 10^{-5}Sr$, we get 
$4.6\times 10^{10}~photons~m^{-2}sec^{-1}Sr^{-1}$. If we ask our signal to be twice this value in 
order to have a SNR =2, we compare $9.6\times 10^{10}~photons~m^{-2}~sec^{-1¨}sr^{-1}$ to our signal:
$0.5\times 10^{-8}~0.5\times 10^{-5}=(extrapolation~from~fig.2)\cdot(bin~width~in~keV)= 
0.25\times 10^{-10}~photons~m^{-2}sec^{-1}sr^{-1}$ which leads to $\lambda/(m_{exch}/100keV)> 2.4\times 10^5$. 
\hfil\hfil\break\noindent
Note that if we use the KECK bin width of $2.2\times 10^{-5}keV$ instead of our bin width of 
$0.5\times 10^{-5}~keV$ and assume the rate to be constant over the whole bin width, our photon rate 
becomes $1.1\times 10^{-10}m^{-2}sec^{-1}Sr^{-1}$ and $\lambda/(m_{exch}/100keV)> 1.7\times 10^5$.

If we increase the observation time to 60 seconds so as to compare with the previous limits, the limits 
are reduced by a factor of 1.67; which brings them to $1.4\times 10^5$ and 
$1\times 10^5$, respectively.

We could also proceed from our previous results and use $2\times 10^{-43}W~m^{-2}Hz^{-1}Sr^{-1}$,  
the bin width of $0.5\times 10^{-5}~keV$ or $1.2\times 10^{12}~Hz$ and the photon energy of 
$1.7\times 10^{-20}J$ to calculate a photon 
rate of about $0.15\times 10^{-10}~photons~m^{-2}sec^{-1}sr^{-1}$. This is quite coherent with the previous result, considering that we extrapolate over 5 orders of magnitude.

This is the ultimate performance of the detector and could be seen as the best case scenario.

Considering [79] and a more recent study by Franceschini and Rodighiero [80] where they model the backgrounds; on figure 6, their model fits nicely 
current data and gives an intensity of $2\times 10^{-9}Wm^{-2}Sr^{-1}$ at 12 microns, which translates 
into $8\times 10^{-23}Wm^{-2}Hz^{-1}Sr^{-1}$ and $\lambda/(m_{exch}/100keV)> 4.5\times 10^5$; which is 
coherent with our previous results obtained from the SNR. 
Here, we have assumed that all the signal comes from the bremsstrahlung process.

Longair [77] gives a signal of about $10^{-22}Wm^{-2}Hz^{-1}Sr^{-1}$ which leads to 
$\lambda/(m_{exch}/100keV)> 4.7\times 10^5$. This is consistent and can be seen as a conservative 
limit as it assumes that all that is measured comes from the bremsstrahlung process.

\noindent 
{\bf Summary}
\hfil\hfil\break\noindent 
We can see then that all reasonable scenarios lead to a constraint of 
$\lambda/(m_{exch}/100keV) > 1-6\times 10^5$; which we summarize by the conservative constraint  
$\lambda/(m_{exch}/100keV) > 5\times 10^5$. As for the proton in the solar wind, we can set a 
relatively conservative bound as $\lambda/(m_{exch}/100MeV) > 14\times 10^5$. 
These results are summarized in Table 4. This means that 
in the process where a DM fermion interacts with a fermion in the solar wind through the exchange of 
a DM scalar particle, this latter particle's mass must be 5-6 orders of magnitude smaller than 
the former's mass. 
Since our process goes like $(\lambda/m_{exch})^4$ we see, from Tables 2 that this limit is fairly 
stable for any DM masses. What is interesting though , is that the contribution of the 
proton in the solar wind probes a very different mass domain in the DM particles than the electron.

Furthermore, one way that could possibly help disentangle this bremsstrahlung process is that it is stronger as we look closer to the sun and is at its weakest when we look opposite the sun. Subtracting 
the signals at two observation angles will certainly reduce the amplitude of the signal, but it could 
also reduce greatly the potential backgrounds.

\noindent{\bf Galactic Emissions \hfil\hfil}
\hfil\hfil\break\noindent
The bremsstrahlung process studied here raises an interesting question regarding its production 
in our galaxy. One would think that any star in the galaxy has its own stellar wind similar to that 
of our Sun, therefore this process should take place for any star. The question is how much of 
the production will reach the Earth. Irrespective of the DM density throughout the galaxy, 
most of the bremsstrahlung production will take place relatively close to the stars, 
where the stellar wind density is largest. As the star density is not very large in our immediate 
galactic neighborhood, and assuming that the DM density is relatively constant in our galactic neighborhood, this process likely would not produce much signal in our neighborhood. When we consider 
the center of the galaxy however, the signal could get substantially larger due to the much larger 
star density. It remains to be seen/calculated how much production will take place as we go farther and 
farther away from the star and how much will reach the Earth. This bremsstrahlung could also 
take place in the collision between high energy electrons or protons/ions and DM particles in the 
cosmos as opposed to the stellar winds. This would add a new source of low energy photons on top of the 
regular bremsstrahlung process known as free-free emission.  
Certainly interesting in itself, such a study is beyond the scope of this analysis.

\noindent{\bf {Cosmology and the 21 cm Line}\hfil\hfil}
\hfil\hfil\break\noindent
The 21 cm line has been studied for a long time and is now used to study the early universe.
[113, 114] The emission from the early hydrogen molecules will be cosmologically 
red-shifted. In the bremsstrahlung process, the spectrum is continuous and extends well 
beyond 21 cm. As we see from our figures, our results can be safely extrapolated by a factor of 10, 
at the very least towards the lower energies. The CHIME collaboration, for example, [115,116] 
studies the 
400-500 MHz band ($60-75~cm$); which corresponds to $(2.4 - 3.33)\times 10^{-6}eV$. The field of view 
of the CHIME detector is large at about 240 square degrees (2 x 120 degrees) which leads to a solid 
angle of about 50 msr. The detector temperature is about 50K. 
This leads to an intensity of approximately $3.3\times 10^{-16}Wm^{-2}Hz^{-1}Sr^{-1}$ 
and $\lambda/(m_{exch}/100keV)>200\times 10^5$. 
We see then that the CHIME detector is not very sensitive to this type of bremsstrahlung emission.
%This process could eventually become a potential background in the search for cosmological 
%hydrogen emission. 

\noindent{\bf {Solar Magnetic Field}\hfil\hfil}
\hfil\hfil\break\noindent
As one sees clearly on figure 2, the orientation that aims closest to the sun is the one that 
leads to the strongest signal. There are two reasons for this. First, when we do not take 
into account the magnetic field of the Sun, the solar wind is at its highest density closest 
to the Sun. Second, it is also in this direction that the solar wind 
will have the largest fraction of head-on collisions, which leads to the highest CME, and 
highest cross-sections. As the magnetic field of the sun changes the direction of the 
solar wind particles, it is likely this orientation that will suffer the largest decrease 
in signal since a fair amount of the head-on collisions will not happen; the density of 
solar wind particle should still remain largest at this orientation however. Similarly, the effects 
of the solar magnetic field should be smaller at 135 degrees and even less so at 45 and 5 degrees. 
The last two orientations, when we have our back to the sun essentially, might even have 
an increase in bremsstrahlung flux since some solar wind particles could hit the 
DM particles at a smaller colliding angle. One could say that, due to the solar magnetic field, 
our results are optimistic at 175 degrees, and more realistic at 135 degrees.

\noindent{\bf {Conclusions}\hfil\hfil}
\hfil\hfil\break\noindent
We have considered the possibility that the solar wind (which we take here as an equal number of protons
and electrons) emits bremsstrahlung radiation when encountering DM particles within the solar system. 
We have considered data from ARCADE2 and Planck and 
concentrated on wavelengths of 21 cm (VLA), 3 mm (ALMA) with an extension to 12 microns (KECK). 
We conclude that in a conservative scenario, the interaction of the protons and electrons of the 
solar wind with a DM particle of 
comparable mass through the exchange of a much lighter messenger (eV for the electron and 
keV for the proton) could lead to an observable signal. We can summarize this limit by 
$\lambda/(m_{exch}/100 keV) > 5 \times 10^5$ for the electron and 
$\lambda/(m_{exch}/100MeV) > 14\times 10^5$ for the proton. The bremsstrahlung signal is stronger 
when the colliding masses are similar, be they $m_{electron}$ or $m_{proton}$ and the variation in 
cross-sections as the DM masses are varied will not alter much the constraints since the 
cross-sections go like $(\lambda/m_{exchange})^4$ to first order.  
Unfortunately, a scenario where the solar wind particles interact directly with the DM particles 
in a Higgs-like fashion does not lead to an observable signal since the cross-section goes like 
$\lambda^4$.

\noindent{\bf {Acknowledgements}\hfil\hfil}
\hfil\hfil\break\noindent
I want to thank my colleague C. Hamzaoui and also G. Hellou for interesting and stimulating discussions. I wish to thank R.Adam, D.J.Fixsen, P.Lubin, N.K. Nikolova and S. Ransom for 
discussions and informations about radio detection; any misunderstanding is entirely mine. 
I want to thank CalculCanada and CalculQu\'ebec without whose computers this work would not have been 
possible.

\hfil\hfil\break
\noindent{\bf {References}\hfil\hfil}
\hfil\hfil\break\noindent
\noindent
{\it 1-}~F.Zwicky,Ap.J.86,217(1937)
\hfil\hfil\break\noindent
%{\it 2-} F.Zwicky, Hel.Ph.Acta, 6; 110-127 (1933)
%\hfil\hfil\break\noindent
{\it 2-}~V.C.Rubin,W.K.J.Ford,Ap.J.159, 379-403
\hfil\hfil\break\noindent
{\it 3-}~V.C.Rubin,W.K.J.Ford,N.Thonnard,1978,
\hfil\hfil\break\noindent\phantom{sss}Ap.J.L,225,L107
\hfil\hfil\break\noindent
{\it 4-}~A.Bosma, 1981, AJ, 86, 1825
\hfil\hfil\break\noindent
{\it 5-}~F.Lelli, S.S.McGaugh, J.M.Schombert; Astron.J. 
\hfil\hfil\break\noindent\phantom{sss}152 (2016) 157 
%\hfil\hfil\break\noindent\phantom{sssss}
e-Print: 1606.09251 [astro-ph.GA]
\hfil\hfil\break\noindent
{\it 6-}~E.Noordermeer, J.M. van der Hulst, R.Sancisi,
\hfil\hfil\break\noindent\phantom{sss}
R.A.Swaters, T.S. van Albada;
\hfil\hfil\break\noindent\phantom{sss}
Astrophys. 442 (2005) 137 
\hfil\hfil\break\noindent\phantom{sss}
• e-Print: astro-ph/0508319 [astro-ph]
\hfil\hfil\break\noindent
{\it 7-}~M.Cirelli, A.Strumia, J.Zupan  2406.01705 [hep-ph]
\hfil\hfil\break\noindent
{\it 8-}~M.Drewes, T. Lasserre, A. Merle, S. Mertens,
\hfil\hfil\break\noindent\phantom{sss}R. Adhikari et al. (Feb 15, 2016)
\hfil\hfil\break\noindent\phantom{sss}
JCAP 01 (2017) 025 • e-Print: 1602.04816 [hep-ph]
\hfil\hfil\break\noindent
{\it 9-}~H. Baer, V.Barger, A.Mustafayev; Phys.Rev.D 85
\hfil\hfil\break\noindent\phantom{sss}(2012) 075010 
e-Print: 1112.3017 [hep-ph]
\hfil\hfil\break\noindent
{\it 10-}~J.Angle XENON10 Collaboration,Phys.Rev.Lett.
\hfil\hfil\break\noindent\phantom{ssss}101 (2008) 091301;
e-Print: 0805.2939 [astro-ph]
\hfil\hfil\break\noindent
{\it 11-}~D.S.Akerib, LUX Collaboration,Phys.Rev.D 102
\hfil\hfil\break\noindent\phantom{ssss}(2020) 11, 112002; 
e-Print: 2004.06304;
\hfil\hfil\break\noindent
{\it 12-}~D.S.Akerib et al. (LUX-ZEPLIN Collaboration),
\hfil\hfil\break\noindent\phantom{ssss}Phys. Rev. D10;
e-print: 052002 (2020);
\hfil\hfil\break\noindent
{\it 13-}~J.Aalbers,LUX-ZEPLIN Collaboration,  eprint:
\hfil\hfil\break\noindent\phantom{ssss}
2410.17036 [hep-ex];
\hfil\hfil\break
{\it 14-}~J.Aalbers,LUX-ZEPLIN Collaboration, Commun.
\hfil\hfil\break\noindent\phantom{ssss}
Phys. 7 (2024) 1, 292; e-Print: 2406.02441 [hep-ex]
\hfil\hfil\break\noindent
{\it 15-}~A.R.Zhitnitsky,JCAP 10 (2003) 010;
\hfil\hfil\break\noindent\phantom{ssss}
e-Print: hep-ph/0202161 [hep-ph]
\hfil\hfil\break\noindent
{\it 16-}~Steven Weinberg, Phys. Rev. Lett. 40, 223 (1978); 
\hfil\hfil\break\noindent
{\it 17-}~Frank Wilczek, Phys. Rev. Lett. 40, 279 (1978);
\hfil\hfil\break\noindent
{\it 18-}~M.Dine, W.Fischler, M.Srednicki, Phys.Lett.B 104
\hfil\hfil\break\noindent\phantom{ssss}
(1981) 199-202
\hfil\hfil\break\noindent
{\it 19-}~C.Arina, E.Del Nobile, P.Panci Nature Phys. 17 
\hfil\hfil\break\noindent\phantom{ssss}(2021) 12, 1396-1401; 
\hfil\hfil\break\noindent\phantom{ssss}
e-Print: 2102.13379 [astro-ph.CO]
\hfil\hfil\break\noindent
{\it 20-}~G.Renzi, J.A.Aguilar-Sanchez, PoS ICRC2023
\hfil\hfil\break\noindent\phantom{ssss}
(2023) 1393; Contribution to: ICRC2023, 1393
\hfil\hfil\break\noindent
{\it 21-}~E.Aprile, et als, XENON100 Collaboration,
\hfil\hfil\break\noindent\phantom{ssss}
Phys.Rev.{\bf D95}, (2016)122001; 
\hfil\hfil\break\noindent
%{\it 22-}~E.Aprile, et als, XENON Collaboration, Eur.Phys.J.{\bf C77},(2017)no.12, 881
%\hfil\hfil\break\noindent
{\it 22-}~C.E.Aalseth, et als, CoGENT Collaboration,
\hfil\hfil\break\noindent\phantom{ssss}
Phys.Rev.{\bf D88}, (2013) 012002;
\hfil\hfil\break\noindent
{\it 23-}~D.S.Akerib, et als, LUX Collaboration,
\hfil\hfil\break\noindent\phantom{ssss}
Phys.Rev.Lett.{\bf 118}, (2017) 021303
\hfil\hfil\break\noindent
{\it 24-}~E.Adams, et als. PICO Collaboration, 
\hfil\hfil\break\noindent\phantom{ssss}
Phys.Rev.D 108 (2023) 6, 062003
\hfil\hfil\break\noindent\phantom{ssss}
e-Print: 2301.08993 [astro-ph.CO]; 
\hfil\hfil\break\noindent
{\it 25-}~S.Archambault et als. PICASSO Collaboration,
\hfil\hfil\break\noindent\phantom{ssss}
Phys.Lett.B 711 (2012) 153-161
\hfil\hfil\break\noindent\phantom{ssss}
e-Print: 1202.1240 [hep-ex]
\hfil\hfil\break\noindent
{\it 26-}~J.Angle, XENON10 Collaboration
\hfil\hfil\break\noindent\phantom{ssss}
Phys.Rev.Lett. 107 (2011) 051301,
\hfil\hfil\break\noindent
{\it 27-}~J.Angle, XENON10 Collaboration
\hfil\hfil\break\noindent\phantom{ssss}
Phys.Rev.Lett. 110 (2013) 249901;
\hfil\hfil\break\noindent\phantom{ssss}
e-Print: 1104.3088 [astro-ph.CO
\hfil\hfil\break\noindent
{\it 28-}~Y.Ema, M. Pospelov, A.Ray JHEP 07 (2024)
\hfil\hfil\break\noindent\phantom{ssss}
094;e-Print: 2402.03431 [hep-ph]
\hfil\hfil\break\noindent
{\it 29-}~H.Abdallah et als HESS Collaboration, Phys.Rev.D
\hfil\hfil\break\noindent\phantom{ssss}
98 (2018) 6, 062005
\hfil\hfil\break\noindent\phantom{ssss}
e-Print: 1807.07113 [astro-ph.CO]
\hfil\hfil\break\noindent
{\it 30-}~H.Abdallah et als. HESS Collaboration,
\hfil\hfil\break\noindent\phantom{ssss}
Phys. Rev. Lett. 129, 111101 (2022),
\hfil\hfil\break\noindent\phantom{ssss}
e-print: 2207.10471v1 [astro-ph.HE]
\hfil\hfil\break\noindent
{\it 31-}~The Atlas Collaboration Sci.Bull. 69 (2024) 19,
\hfil\hfil\break\noindent\phantom{ssss}
3005-3035; e-Print: 2306.00641 [hep-ex];
\hfil\hfil\break\noindent
{\it 32-}~M.Felcini, Atlas and CMS Collaborations
\hfil\hfil\break\noindent\phantom{ssss}
2409.19216 [hep-ex]
\hfil\hfil\break\noindent
{\it 33-}~D.P. Adan Atlas and CMS Collaborations,
\hfil\hfil\break\noindent\phantom{ssss}
Moriond 2022 EW;
e-Print: 2301.10141 [hep-ex]
\hfil\hfil\break\noindent
{\it 34-}~S.Afach et als. GNOME Collaboration;
\hfil\hfil\break\noindent\phantom{ssss}
Nature Phys. 17 (2021) 12, 1396-1401
\hfil\hfil\break\noindent\phantom{ssss}
e-Print: 2102.13379 [astro-ph.CO]
\hfil\hfil\break\noindent
{\it 35-}~S.Afach GNOME Collaboration;
\hfil\hfil\break\noindent\phantom{ssss}
Annalen Phys. 536 (2024) 1, 2300083 
\hfil\hfil\break\noindent
{\it 36-}~R.Bernabei, et als, DAMA/LIBRA Collaboration
\hfil\hfil\break\noindent\phantom{ssss}
Nucl.Phys.Atom.Energy 19 
\hfil\hfil\break\noindent\phantom{ssss}
(2018) 4, 307-325• e-Print: 1805.10486 [hep-ex]; 
\hfil\hfil\break\noindent
{\it 37-}~R.Bernabei et als DAMA/LIBRA Collaboration
\hfil\hfil\break\noindent\phantom{ssss}
Eur.Phys.J.C 73 (2013) 2648
\hfil\hfil\break\noindent\phantom{ssss}
e-Print: 1308.5109 [astro-ph.GA]; 
\hfil\hfil\break\noindent
{\it 38-}~G.Adhikari et als. COSINE-100 Collaboration
\hfil\hfil\break\noindent\phantom{ssss}
Phys.Rev.D 106 (2022) 5, 052005
\hfil\hfil\break\noindent\phantom{ssss}
e-Print: 2111.08863 [hep-ex];
\hfil\hfil\break\noindent
{\it 39-}~D.S.Akerib et als. LUX Collaboration,
\hfil\hfil\break\noindent\phantom{ssss}
Phys.Rev.D 98 (2018) 6, 062005
\hfil\hfil\break\noindent\phantom{ssss}
e-Print: 1807.07113 [astro-ph.CO]
\hfil\hfil\break\noindent
{\it 40-}~E.Aprile et als. XENON1T Collaboration
\hfil\hfil\break\noindent\phantom{ssss}
Phys.Rev.D 102 (2020) 7, 072004
\hfil\hfil\break\noindent\phantom{ssss}
e-Print: 2006.09721 [hep-ex]; 
\hfil\hfil\break\noindent
{\it 41-}~E.Aprile et alx. XENONnT Collaboration
\hfil\hfil\break\noindent\phantom{ssss}
Eur.Phys.J.C 84 (2024) 2, 138
\hfil\hfil\break\noindent\phantom{ssss}
e-Print: 2309.11996 [hep-ex].
\hfil\hfil\break\noindent
{\it 42-}~E.Aprile et als, XENONnT Collaboration
\hfil\hfil\break\noindent\phantom{ssss}
Physical Review Letters. 129 (16): 
\hfil\hfil\break\noindent\phantom{ssss}
161805; e-print: 2207.11330 [hep-ex]
\hfil\hfil\break\noindent
{\it 43-}~V.Springel et als., Nature, 435:629-636 (2005)
\hfil\hfil\break\noindent\phantom{ssss}
astro-ph/0504097;
\hfil\hfil\break\noindent
{\it 44-}~R.E.Smith et als. VIRGO Consortium,
\hfil\hfil\break\noindent\phantom{ssss}
Mon.Not.Roy.Astron.Soc. 341 (2003) 1311
\hfil\hfil\break\noindent\phantom{ssss}
e-Print: astro-ph/0207664 [astro-ph]
\hfil\hfil\break\noindent
{\it 45-}~Planck collaboration, Planck 2015 results. XIV. 
\hfil\hfil\break\noindent\phantom{ssss}
Dark energy and modified gravity, 
\hfil\hfil\break\noindent\phantom{ssss}
Astron.Astrophys. 594 (2016) A14 [1502.01590].
\hfil\hfil\break\noindent
{\it 46-}~N.Aghanim, PLANCK Collaboration
\hfil\hfil\break\noindent\phantom{ssss}
Astron.Astrophys.641(2020)A1
\hfil\hfil\break\noindent\phantom{ssss}
e-Print: 1807.06205 [astro-ph.CO]
\hfil\hfil\break\noindent
{\it 47-}~T.M.C.Abbott DES collaboration,
\hfil\hfil\break\noindent\phantom{ssss}
%Dark Energy Survey Year 3 results: 
%\hfil\hfil\break\noindent\phantom{ssss}
%Constraints on extensions to ΛCDM with weak lensing and galaxy clustering, 
%\hfil\hfil\break\noindent\phantom{ssss}
Phys. Rev. D 107 (2023) 083504 [2207.05766].
\hfil\hfil\break\noindent
%{\it 48-}~Making the Case for Conformal Gravity, Philip D. Mannheim (Jan, 2011)
{\it 48-}P.D.Mannheim, Found.Phys. 42 (2012) 388-420
\hfil\hfil\break\noindent\phantom{ssss}
e-Print: 1101.2186 [hep-th]
\hfil\hfil\break\noindent
{\it 49-}P.D.Mannheim, J.G.O'Brien,Phys.Rev.D 85 (2012)
\hfil\hfil\break\noindent\phantom{ssss}
124020; e-Print: 1011.3495 [astro-ph.CO]
\hfil\hfil\break\noindent
{\it 50-}~M. Milgrom Astrophys.J. 270 (1983) 365-370,
\hfil\hfil\break\noindent\phantom{ssss}371-383, 384-389
\hfil\hfil\break\noindent
{\it 51-}~T.Mistele, S. McGaugh, F.Lelli, J.Schombert, P.Li,
\hfil\hfil\break\noindent\phantom{ssss}
Astrophys.J.Lett. 969 (2024) 1, L3;
\hfil\hfil\break\noindent\phantom{ssss}
e-Print: 2406.09685 [astro-ph.GA]
\hfil\hfil\break\noindent
%{\it 23- 21cm} This reference is important but stops compiling
{\it 52-}~G.Couture,Phys.Rev.D 105(2022)5,055003;
\hfil\hfil\break\noindent\phantom{ssss}
e-Print:2106.16027 [hep-ph]
\hfil\hfil\break\noindent
{\it 53-}~S.Furlanetto, S.Peng Oh, F.Briggs;
\hfil\hfil\break\noindent\phantom{ssss}
Phys.Rept.433(2006)181-301; 
\hfil\hfil\break\noindent\phantom{ssss}
e-Print: 0608032 [astro-ph];
\hfil\hfil\break\noindent
{\it 54-}~H.Bethe, W.Heitler, Proc.Roy.Soc.{\bf 146}(1934)83
\hfil\hfil\break\noindent
{\it 55-}~{\it Classical Electrodynamics}, J.D. Jackson;
\hfil\hfil\break\noindent\phantom{ssss}
Wiley, New York, $2^{nd}$ Edition, 1975
\hfil\hfil\break\noindent
{\it 56-}~I.B. Khriplovich, e-Print: 1005.1778 [astro-ph.EP];
\hfil\hfil\break\noindent
{\it 57-}~J.Edsjo,A.H.G.Peter,e-Print:1004.5258[astro-ph.EP]
\hfil\hfil\break\noindent
{\it 58-}~A. Gould, ApJ 368 (1991) 610;
\hfil\hfil\break\noindent
{\it 59-}~O. Gron, H.H.Soleng  Astrophys.J. 456 (1996)
\hfil\hfil\break\noindent\phantom{ssss}
445-448; e-Print: 9507051 [astro-ph]; 
\hfil\hfil\break\noindent
{\it 60-}~F.Munyaneza,R.D.Viollier
\hfil\hfil\break\noindent\phantom{ssss}
e-Print:astro-ph/9910566[astro-ph];
\hfil\hfil\break\noindent
{\it 61-}~I.Lopes, J.Silk, Phys.Rev.D 99 (2019) 2, 023008;
\hfil\hfil\break\noindent\phantom{ssss}
e-Print: 1812.07426 [hep-ph];
\hfil\hfil\break\noindent
{\it 62-}~I.Lopes, J.Silk Astrophys.J.Lett. 752 (2012) L129; 
\hfil\hfil\break\noindent\phantom{ssss}e-Print: 1309.7573 [astro-ph.SR];
\hfil\hfil\break\noindent
{\it 63-}~M.Taoso, F.Iocco, G.Meynet, G.Bertone,
\hfil\hfil\break\noindent\phantom{ssss}
P.Eggenberger, Phys.Rev.D 82(2010) 083509;
\hfil\hfil\break\noindent\phantom{ssss}
e-Print: 1005.5711 [astro-ph.CO]
\hfil\hfil\break\noindent
{\it 64-}~M.Klasen,M.Pohl,G.Sigi,Prog.in Part.and
\hfil\hfil\break\noindent\phantom{ssss}
Nucl.Phys.85 (2015) 1-32;hep-ph:1507.03800
\hfil\hfil\break\noindent
{\it 65-}~J.I.Read, J.Phys.{\bf G41}(2014)063101;
\hfil\hfil\break\noindent\phantom{ssss}
astro-ph.GA: 1404.1938
\hfil\hfil\break\noindent
{\it 66-}~C.J.Copi,L.M.Krauss,Phys.Rev.{\bf D63}(2001),043507;
\hfil\hfil\break\noindent\phantom{ssss}
astro-ph:0009467
\hfil\hfil\break\noindent
{\it 67-}~P.Salucci, F.Nesti, G.Gentile, C.F.Martins,
\hfil\hfil\break\noindent\phantom{ssss}
Astron.Astrophysics,{\bf 523}(2010)A83
\hfil\hfil\break\noindent
{\it 68-}~{\it Relativistic Quantum Mechanics}, Bjorken J.D. and
\hfil\hfil\break\noindent\phantom{ssss}
S.D. Drell; Vol. 1, McGraw-Hill,New York, 1964
\hfil\hfil\break\noindent
{\it 69-}~{\it An Introduction to Quantum Field Theory}, 
\hfil\hfil\break\noindent\phantom{ssss}
Peskin, M.E. and D. V. Schroeder;
\hfil\hfil\break\noindent\phantom{ssss}
Addison Wesley, New York, 1995
\hfil\hfil\break\noindent
{\it 70}~{\it Relativistic Quantum Mechanics and Field Theory},
\hfil\hfil\break\noindent\phantom{ssss}
Gross, F.,Wiley Interscience, New York, 1993
\hfil\hfil\break\noindent
{\it 71}~{\it Relativistic Kinematics}, R. Hagedorn;
\hfil\hfil\break\noindent\phantom{ssss}
Benjamin/Cummings, London, 1963
\hfil\hfil\break\noindent
{\it 72-}~{\it Collider Physics}, Barger, V.D. and R.J.N. Phillips,
\hfil\hfil\break\noindent\phantom{ssss}
Addison Wesley, New York, 1987
\hfil\hfil\break\noindent
{\it 73-}~A.Savitzky,M.J.E.Golay,Analyt.Chem.{\bf 8}
\hfil\hfil\break\noindent\phantom{ssss}
(1964)1627-1639
\hfil\hfil\break\noindent
{\it 74-}~W.Herr, B.Muratori,{\it CERN Acc. School and
\hfil\hfil\break\noindent\phantom{ssss}
DESY Zeuthen: Accelerator Physics},(2003)361-377
\hfil\hfil\break\noindent
{\it 75-}~H.Burkhardt, P. Grafstrom, LHC Project Report
\hfil\hfil\break\noindent\phantom{ssss}
1019 (2007)
\hfil\hfil\break\noindent
{\it 76-}~G.Couture, Eu.J.P., (2012)33,3,479
\hfil\hfil\break\noindent
{\it 77-}~M.S.Longair, "The Radio Background Emission...", 
\hfil\hfil\break\noindent\phantom{ssss}
in Extragalactic Background Radiation", 
\hfil\hfil\break\noindent\phantom{ssss}
Space Telescope Science Institute Symposium, 
\hfil\hfil\break\noindent\phantom{ssss}
Series 7, 1995, Eds. D. Calzetti, M.Livio, P.Madau
\hfil\hfil\break\noindent 
{\it 78-}~J.J.Condon, ASP Conference Series,{\bf 278}, 2002, 
\hfil\hfil\break\noindent\phantom{ssss}
S. Stanimirovic, D.R.Altschuler, P.F. Goldsmith,
\hfil\hfil\break\noindent\phantom{ssss}
C.J. Salter, Eds., p.155 (2002)
\hfil\hfil\break\noindent
{\it 79-}~A.Cooray, Royal Society Open Science, 
\hfil\hfil\break\noindent\phantom{ssss}
e-print: 1602:03512 [astro-ph.CO]
\hfil\hfil\break\noindent
{\it 80-}~A.Franceschini, G. Rodighiero, A\&A 603, {\bf A34}, 
\hfil\hfil\break\noindent\phantom{ssss}
(2017); e-Print:1705.10256 [astro-ph.HE]
\hfil\hfil\break\noindent
{\it 81-}~J.L. Weiland, Charles L. Bennett, Graeme E. 
\hfil\hfil\break\noindent\phantom{ssss}
Addison, Mark Halpern, Gary Hinshaw 
\hfil\hfil\break\noindent\phantom{ssss}
(2024)Astrophys.J. 984 (2025) 1, 80 • 
\hfil\hfil\break\noindent\phantom{ssss}
e-Print: 2409.13132 [astro-ph.CO]
\hfil\hfil\break\noindent
{\it 82-}~J.J.Condon, W.D.Cotton, E.B.Fomalont,
\hfil\hfil\break\noindent\phantom{ssss}
K.I.Kellermann, N.Miller, R.A.Perly, D.Scott, 
\hfil\hfil\break\noindent\phantom{ssss}
T.Vernstrom, J.D.Wall (2012) Astrophys.J. 758 
\hfil\hfil\break\noindent\phantom{ssss}
(2012) 23; e-Print: 1207.2439 [astro-ph.CO]
\hfil\hfil\break\noindent
{\it 83-}~"Essential Radio Astronomy", S. Ransom, 
\hfil\hfil\break\noindent\phantom{ssss}
https://www.cv.nrao.edu/~sransom/web/
\hfil\hfil\break\noindent 
{\it 84-}~Basics of Radio Astronomy, T. Wilson, NRAO 
\hfil\hfil\break\noindent\phantom{ssss}
Synthesis Workshop, 2014 (pdf)
\hfil\hfil\break\noindent 
{\it 85-}~P. Platania, M.Bensadoun, M.Bersanelli, 
\hfil\hfil\break\noindent\phantom{ssss}
G.DeAmici, A.Kogut, S.Levin, D. Maino,
\hfil\hfil\break\noindent\phantom{ssss}
G.F. Smoot(1997)Astrophys.J. 505 (1998) 473
\hfil\hfil\break\noindent\phantom{ssss}
e-Print: astro-ph/9707252 [astro-ph]
\hfil\hfil\break\noindent 
{\it 86-}~R.D.Davies, A.Wilkinson (Apr, 1998) Contribution 
\hfil\hfil\break\noindent\phantom{ssss}
to 33rd Rencontres de Moriond: Fundamental 
\hfil\hfil\break\noindent\phantom{ssss}
Parameters in Cosmology, 175-182 • 
\hfil\hfil\break\noindent\phantom{ssss}
e-Print: astro-ph/9804208 [astro-ph]
\hfil\hfil\break\noindent 
{\it 87-}~Measuring and calibrating Galactic synchrotron 
\hfil\hfil\break\noindent\phantom{ssss}
emission, W.Reich, P.Reich (2008)
\hfil\hfil\break\noindent\phantom{ssss}
Contribution to: IAU Symposium 259, 603-612 
\hfil\hfil\break\noindent\phantom{ssss}
e-Print: 0812.4128 [astro-ph]
\hfil\hfil\break\noindent 
{\it 88-}~C.Dickinson, R.D.Davies, R.J.Davis (Feb, 2003)
\hfil\hfil\break\noindent\phantom{ssss}
Mon.Not.Roy.Astron.Soc. 341 (2003) 369 • 
\hfil\hfil\break\noindent\phantom{ssss}
e-Print: astro-ph/0302024 [astro-ph]
\hfil\hfil\break\noindent 
{\it 89-}~R.Galvan-Madrid,D.J.Diaz-Gonzalez,F.Motte 
\hfil\hfil\break\noindent\phantom{ssss}
{\it et als}, e-Print: 2407.07359 [astro-ph.GA]
\hfil\hfil\break\noindent 
{\it 90-}~D.Kandel, A.Lazarian, D.Pogosyan (2017) 
\hfil\hfil\break\noindent\phantom{ssss}
Mon.Not.Roy.Astron.Soc. 478 (2018) 1, 530-540; 
\hfil\hfil\break\noindent\phantom{ssss}
e-Print: 1711.03161 [astro-ph.GA]
\hfil\hfil\break\noindent 
{\it 91-}~N.Fornengo, R.A.Lineros, M.Regis, M.Taoso 
\hfil\hfil\break\noindent\phantom{ssss}
(2014) JCAP 04 (2014) 008 
\hfil\hfil\break\noindent\phantom{ssss}
e-Print: 1402.2218 [astro-ph.CO]
\hfil\hfil\break\noindent 
{\it 92-}~R.A.Battye, B.Garbrecht, J.I. McDonald, F.Pace,
\hfil\hfil\break\noindent\phantom{ssss}
S.Srinivasan (Oct 25, 2019) Phys.Rev.D 102 (2020)
\hfil\hfil\break\noindent\phantom{ssss}
2, 023504 • e-Print: 1910.11907 [astro-ph.CO]
\hfil\hfil\break\noindent
{\it 93-}~R.Barkana, A.Fialkov, H. Liu, N.J. Outmezguine
\hfil\hfil\break\noindent\phantom{ssss}
(2022) Phys.Rev.D 108 (2023) 6, 063503 
\hfil\hfil\break\noindent\phantom{ssss}
e-Print: 2212.08082 [hep-ph]
\hfil\hfil\break\noindent 
{\it 94-}~D.J. Fixsen, A. Kogut, S. Levin, M. Limon,
\hfil\hfil\break\noindent\phantom{ssss}
P. Lubin, P.Mirel, M.Seiffert, J.Singal, E.Wollack,
\hfil\hfil\break\noindent\phantom{ssss}
T.Villela, C.A.Wuensche (2009) Astrophys.J. 734
\hfil\hfil\break\noindent\phantom{ssss}
(2011) 5; e-Print: 0901.0555 [astro-ph.CO]
\hfil\hfil\break\noindent
{\it 95-}~M. Seiffert, D.J. Fixsen, A. Kogut, S.M. Levin,
\hfil\hfil\break\noindent\phantom{ssss}
M.Limon, P.M.Lubin, P.Mirel, J.Singal, T.Villela,
\hfil\hfil\break\noindent\phantom{ssss}
E.Wollack, C.A.Wuensche (2009); 
\hfil\hfil\break\noindent\phantom{ssss}
e-Print: 0901.0559 [astro-ph.CO]
\hfil\hfil\break\noindent
{\it 96-}~A.Addazi, S.Capozziello, Q.Gan, G.Lambiase,
\hfil\hfil\break\noindent\phantom{ssss}
R.Samanta(2024)e-Print:2411.09042 [astro-ph.CO]
\hfil\hfil\break\noindent 
{\it 97-}~S.K.Acharya, B.Cyr, J.Chluba (2023)
\hfil\hfil\break\noindent\phantom{ssss}
Mon.Not.Roy.Astron.Soc. 523 (2023) 2, 1908-1918 
\hfil\hfil\break\noindent\phantom{ssss}
e-Print: 2303.17311 [astro-ph.CO]
\hfil\hfil\break\noindent
{\it 98-}~N.Fornengo, R.Lineros, M.Regis, M.Taoso 
\hfil\hfil\break\noindent\phantom{ssss}
(2011) Phys.Rev.Lett. 107 (2011) 271302 
\hfil\hfil\break\noindent\phantom{ssss}
e-Print: 1108.0569 [hep-ph]
\hfil\hfil\break\noindent 
{\it 99-}~D.J.Fixsen, APJ, {\bf 707}, (2009), 916-920;
\hfil\hfil\break\noindent\phantom{ssss}
e-Print: 0911.1955 [astro-ph.CO]
\hfil\hfil\break\noindent
{\it 100-}~J.D.Bowman, A.E.E. Rogers, R.A.Monsalve,
\hfil\hfil\break\noindent\phantom{ssss}
T.J.Mozdzen, N.Mahesh(2018) Nature 555, 7694,
\hfil\hfil\break\noindent\phantom{ssss}
67-70; e-Print: 1810.05912 [astro-ph.CO]
\hfil\hfil\break\noindent
{\it 101-}~G.D'Amico, P.Panci, A.Strumia (2018)
\hfil\hfil\break\noindent\phantom{ssss}
Phys.Rev.Lett. 121 (2018) 1, 011103 
\hfil\hfil\break\noindent\phantom{ssss}
e-Print: 1803.03629 [astro-ph.CO]
\hfil\hfil\break\noindent
{\it 102-}~J.Dowell, G.B.Taylor (2018) Astrophys.J.Lett. 
\hfil\hfil\break\noindent\phantom{ssss}
858 (2018) 1, L9 
\hfil\hfil\break\noindent\phantom{ssss}
e-Print: 1804.08581 [astro-ph.CO]
\hfil\hfil\break\noindent
{\it 103-}~N.Mahesh, J.D.Bowman, T.J.Mozdzen. A.E.E.
\hfil\hfil\break\noindent\phantom{ssss}
Rogers, R.A.Monsalve et al. (2021)Astron.J. 
\hfil\hfil\break\noindent\phantom{ssss}
162 (2021) 2, 38 • e-Print: 2103.00423 [astro-ph.IM]
\hfil\hfil\break\noindent
{\it 104-}~S.Singh, Jishnu Nambissan T., R.Subrahmanyan,
\hfil\hfil\break\noindent\phantom{ssss}
N.U.Shankar, B.S.Girish (Dec 13, 2021) Nature
\hfil\hfil\break\noindent\phantom{ssss}
Astron. 6 (2022) 5, 607-617 
\hfil\hfil\break\noindent\phantom{ssss}
e-Print: 2112.06778 [astro-ph.CO]
\hfil\hfil\break\noindent
{\it 105-}~PLANCK Collaboration, R.Adam {\it et als},
\hfil\hfil\break\noindent\phantom{ssss}
Astron.Astrophys. 594 (2016) A10 
\hfil\hfil\break\noindent\phantom{ssss}
e-Print: 1502.01588 [astro-ph.CO]
\hfil\hfil\break\noindent 
{\it 106-}~T.Hunter, C.Brogan; afe.nrao.edu/wiki/pub
\hfil\hfil\break\noindent\phantom{ssss}
/Main/UsefulFormulas/planck.pdf
\hfil\hfil\break\noindent   
{\it 107-}~science.nrao.edu/facilities/vla/docs/manuals/oss/
\hfil\hfil\break\noindent\phantom{ssss}
performance/sensitivity
\hfil\hfil\break\noindent
{\it 108-}~science.nrao.edu/facilities/vla/proposing/TBconv
\hfil\hfil\break\noindent
{\it 109-}~eso.org/public/teles-instr/alma/receiver-bands/
\hfil\hfil\break\noindent
{\it 110-}almascience.eso.org/documents-and-tools
\hfil\hfil\break\noindent\phantom{ssss}
/cycle12/alma-science-primer.pdf
\hfil\hfil\break\noindent
{\it 111-}almascience.eso.org/documents-and-tools/cycle12
\hfil\hfil\break\noindent\phantom{ssss}
/alma-technical-handbook.pdf
\hfil\hfil\break\noindent 
{\it 112-}www2.keck.hawaii.edu/observing/kecktelgde
\hfil\hfil\break\noindent\phantom{ssss}
/ktelinstupdate.pdf
\hfil\hfil\break\noindent 
{\it 113-}~J.R.Pritchard, A. Loeb; Rept.Prog.Phys.75(2012)
\hfil\hfil\hfil\break\noindent\phantom{ssss}
086901; e-Print: 1109.6012 [astro-ph.CO]\hfil\hfil\break\noindent 
{\it 114-}~B.Ciardi, P.Madau (Mar, 2003) Astrophys.J. 596
\hfil\hfil\break\noindent\phantom{ssss}
(2003) 1-8 • e-Print: astro-ph/0303249 [astro-ph]
\hfil\hfil\break\noindent
{\it 115}~M.Amiri, et als, CHIME Collaboration,
\hfil\hfil\break\noindent\phantom{ssss}
Astrophys.J.Supp. 261 (2022) 2, 29 
\hfil\hfil\break\noindent\phantom{sssss}
e-Print: 2201.07869 [astro-ph.IM]
\hfil\hfil\break\noindent
{\it 116}~M.Amiri, et als, CHIME Collaboration,
\hfil\hfil\break\noindent\phantom{ssss}
Astrophys.J. 963 (2024) 1, 23
\hfil\hfil\break\noindent\phantom{sssss}
e-Print: 2309.04404 [astro-ph.CO]
\hfil\hfil\break\noindent
\hfil\hfil\break
\vfil\vfil\eject
\begin{figure}
{\hsize=3.0truein
    \centering
    \includegraphics[scale=0.5]{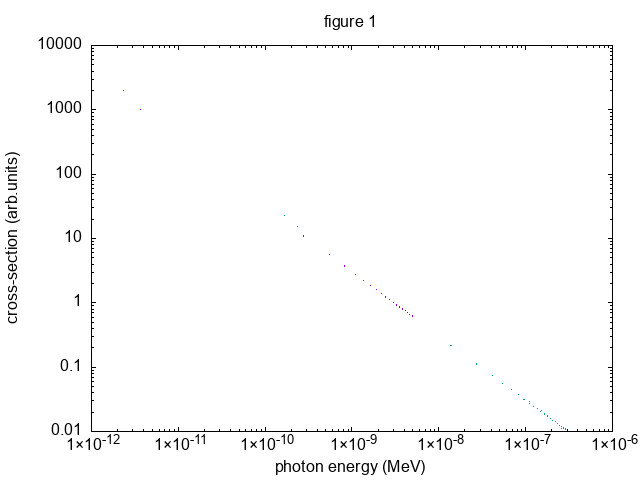}}
\end{figure}
\begin{figure}
{\hsize=3.0truein
    \centering
    \includegraphics[scale=0.5]{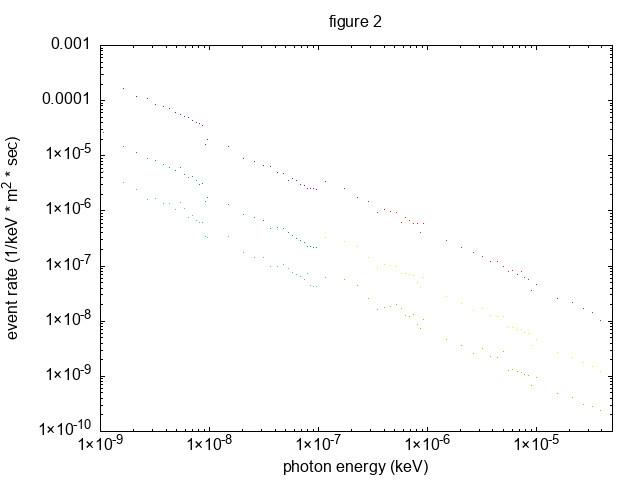}}
\end{figure}
%\end{document}
\vfil\vfil\eject
{\hsize=4.0truein
$$
\vbox{\offinterlineskip
\halign{&\vrule#&\strut\ #\ \cr
\multispan{5}\hfil\bf Table 1\hfil\cr
\hfil\cr
\noalign{\smallskip}
\noalign{\hrule}
height3pt&\omit&&\omit&\cr
&\hfil $M_{_{DM}}~(keV)$ \hfil&&\hfil $\sigma~(arb.units)$\hfil&\cr
height3pt&\omit&&\omit&\cr
\noalign{\hrule}
height3pt&\omit&&\omit&\cr
&\hfil 1000\hfil&&\hfil 4482\hfil&\cr
height3pt&\omit&&\omit&\cr
\noalign{\hrule}
height3pt&\omit&&\omit&\cr
&\hfil 1022\hfil&&\hfil 4845\hfil&\cr
height3pt&\omit&&\omit&\cr
\noalign{\hrule}
height3pt&\omit&&\omit&\cr
&\hfil 1022.0015\hfil&&\hfil $6\times 10^6$\hfil&\cr
height3pt&\omit&&\omit&\cr
\noalign{\hrule}
height3pt&\omit&&\omit&\cr
&\hfil 1022.0018\hfil&&\hfil $3\times 10^7$\hfil&\cr
height3pt&\omit&&\omit&\cr
\noalign{\hrule}
height3pt&\omit&&\omit&\cr
&\hfil 1022.0019\hfil&&\hfil $2\times 10^8$\hfil&\cr
height3pt&\omit&&\omit&\cr
\noalign{\hrule}
height3pt&\omit&&\omit&\cr
&\hfil 1022.0020\hfil&&\hfil $4\times 10^{10}$\hfil&\cr
height3pt&\omit&&\omit&\cr
\noalign{\hrule}
height3pt&\omit&&\omit&\cr
&\hfil 1022.0021\hfil&&\hfil $4\times 10^8$\hfil&\cr
height3pt&\omit&&\omit&\cr
\noalign{\hrule}
height3pt&\omit&&\omit&\cr
&\hfil 1022.0025\hfil&&\hfil $9\times 10^7$\hfil&\cr
height3pt&\omit&&\omit&\cr
\noalign{\hrule}
height3pt&\omit&&\omit&\cr
&\hfil 1023\hfil&&\hfil 4425\hfil&\cr
height3pt&\omit&&\omit&\cr
\noalign{\hrule}
\noalign{\hrule}\noalign{\smallskip}}}
$$}
\vfil\vfil\eject
{\hsize=3.0truein
$$
\vbox{\offinterlineskip
\halign{&\vrule#&\strut\ #\ \cr
\multispan{19}\hfil\bf Table 2A\hfil\cr
\hfil\cr
\noalign{\smallskip}
\noalign{\hrule}
&\omit&&
\multispan{3}\hfil$\beta_2=1\times 10^{-3}$\hfil&&
\multispan{3}\hfil$\beta_2=2\times 10^{-3}$\hfil&&
\multispan{3}\hfil$\beta_2=3\times 10^{-3}$\hfil&&
\multispan{3}\hfil$\beta_2=6\times 10^{-3}$\hfil&\cr
\noalign{\hrule}
height3pt&\omit&&\omit&&\omit&&\omit&&\omit&&\omit&&\omit&&\omit&&\omit&\cr
&\hfil$ M_{_{DM}}~(keV)$\hfil&&\hfil $R_\sigma$\hfil&&\hfil $R_{\rho\sigma}$\hfil&&
\hfil $R_\sigma$\hfil&&\hfil$R_{\rho\sigma}$\hfil&&\hfil$R_\sigma$\hfil&&\hfil $R_{\rho\sigma}$\hfil&&
\hfil $R_\sigma$\hfil&&\hfil$R_{\rho\sigma}$\hfil&\cr
height3pt&\omit&&\omit&&\omit&&\omit&&\omit&&\omit&&\omit&&\omit&&\omit&\cr
\noalign{\hrule}
height3pt&\omit&&\omit&&\omit&&\omit&&\omit&&\omit&&\omit&&\omit&&\omit&\cr
&\hfil 5\hfil&&\hfil 0.49\hfil&&\hfil 0.97\hfil&&\hfil 1.25\hfil&&\hfil 2.5\hfil&&
\hfil 2.6\hfil&&\hfil 5.2\hfil&&\hfil 12.5 \hfil&&\hfil 25.0\hfil&\cr
\noalign{\hrule}
height3pt&\omit&&\omit&&\omit&&\omit&&\omit&&\omit&&\omit&&\omit&&\omit&\cr
&\hfil 10\hfil&&\hfil 1.00\hfil&&\hfil 1.00\hfil&&\hfil 2.6\hfil&&\hfil 2.6\hfil&&
\hfil 5.1\hfil&&\hfil 5.1\hfil&&\hfil 24.9 \hfil&&\hfil 24.9\hfil&\cr
\noalign{\hrule}
height3pt&\omit&&\omit&&\omit&&\omit&&\omit&&\omit&&\omit&&\omit&&\omit&\cr
&\hfil 50\hfil&&\hfil 4.0\hfil&&\hfil 0.79\hfil&&\hfil 11.2\hfil&&\hfil 2.2\hfil&&
\hfil 23.2\hfil&&\hfil 4.6\hfil&&\hfil 116 \hfil&&\hfil 23.2\hfil&\cr
\noalign{\hrule}
height3pt&\omit&&\omit&&\omit&&\omit&&\omit&&\omit&&\omit&&\omit&&\omit&\cr
&\hfil 100\hfil&&\hfil 6.0\hfil&&\hfil 0.6\hfil&&\hfil 16.4\hfil&&\hfil 1.64\hfil&&
\hfil 32.0\hfil&&\hfil 3.2\hfil&&\hfil 139 \hfil&&\hfil 13.9\hfil&\cr
\noalign{\hrule}
height3pt&\omit&&\omit&&\omit&&\omit&&\omit&&\omit&&\omit&&\omit&&\omit&\cr
&\hfil 200\hfil&&\hfil 5.6\hfil&&\hfil 0.28\hfil&&\hfil 14.0\hfil&&\hfil 0.70\hfil&&
\hfil 27.3\hfil&&\hfil 1.37\hfil&&\hfil 118 \hfil&&\hfil 5.9\hfil&\cr
\noalign{\hrule}
height3pt&\omit&&\omit&&\omit&&\omit&&\omit&&\omit&&\omit&&\omit&&\omit&\cr
&\hfil 300\hfil&&\hfil 3.8\hfil&&\hfil 0.12\hfil&&\hfil 9.5\hfil&&\hfil 0.32\hfil&&
\hfil 13.0\hfil&&\hfil 0.43\hfil&&\hfil 80.4 \hfil&&\hfil 2.7\hfil&\cr
\noalign{\hrule}
height3pt&\omit&&\omit&&\omit&&\omit&&\omit&&\omit&&\omit&&\omit&&\omit&\cr
&\hfil 500\hfil&&\hfil 1.7\hfil&&\hfil 0.04\hfil&&\hfil 4.2\hfil&&\hfil 0.08\hfil&&
\hfil 8.7\hfil&&\hfil 0.17\hfil&&\hfil 35.7 \hfil&&\hfil 0.7\hfil&\cr
\noalign{\hrule}
height3pt&\omit&&\omit&&\omit&&\omit&&\omit&&\omit&&\omit&&\omit&&\omit&\cr
&\hfil 1000\hfil&&\hfil 0.32\hfil&&\hfil 0.003\hfil&&\hfil 1.3\hfil&&\hfil 0.013\hfil&&
\hfil 1.56\hfil&&\hfil 0.016\hfil&&\hfil 6.4 \hfil&&\hfil 0.064\hfil&\cr
\noalign{\hrule}
\noalign{\hrule}\noalign{\smallskip}
\multispan{19}Higgs-like scenario for the DM\hfil\hfil\cr
\multispan{19}$M=511keV=M_{exch};\beta_1 = 2\times 10^{-3};\theta_q = 0~(head-on~coll.(HOC))$\hfil\hfil\cr
\multispan{19}$E_\gamma^{min}=5\times 10^{-10}keV;M_\gamma = 1\times 10^{-13}keV; 
RMAX=2.0\times 10^{-3}$\hfil\cr
\multispan{19}For 511-100-511: $M_\gamma = 10^{-13}keV\to \sigma = 75921~ar.units~and$ \hfil\cr
\multispan{19}\phantom{For 511-100-511: }$M_\gamma = 10^{-14}keV\to \sigma = 75923~ar.units$\hfil\cr}}
$$}
{\hsize=4.0truein
$$
\vbox{\offinterlineskip
\halign{&\vrule#&\strut\ #\ \cr
\multispan{19}\hfil\bf Table 2B\hfil\cr
\hfil\cr
\noalign{\smallskip}
\noalign{\hrule}
&\omit&&
\multispan{3}\hfil$\beta_2=1\times 10^{-3}$\hfil&&
\multispan{3}\hfil$\beta_2=2\times 10^{-3}$\hfil&&
\multispan{3}\hfil$\beta_2=3\times 10^{-3}$\hfil&&
\multispan{3}\hfil$\beta_2=6\times 10^{-3}$\hfil&\cr
\noalign{\hrule}
height3pt&\omit&&\omit&&\omit&&\omit&&\omit&&\omit&&\omit&&\omit&&\omit&\cr
&\hfil $M_{_{DM}}~(MeV)$\hfil&&\hfil $R_\sigma$\hfil&&\hfil$R_{\rho\sigma}$\hfil&&
\hfil$R_\sigma$\hfil&&\hfil$R_{\rho\sigma}$\hfil&&\hfil$R_\sigma$\hfil&&\hfil $R_{\rho\sigma}$\hfil&&
\hfil$R_\sigma$\hfil&&\hfil $R_{\rho\sigma}$\hfil&\cr
height3pt&\omit&&\omit&&\omit&&\omit&&\omit&&\omit&&\omit&&\omit&&\omit&\cr
\noalign{\hrule}
height3pt&\omit&&\omit&&\omit&&\omit&&\omit&&\omit&&\omit&&\omit&&\omit&\cr
&\hfil 5\hfil&&\hfil 0.44\hfil&&\hfil 0.88\hfil&&\hfil 1.22\hfil&&\hfil 2.44\hfil&&
\hfil 2.52\hfil&&\hfil 5.0\hfil&&\hfil 11.8 \hfil&&\hfil 24\hfil&\cr
\noalign{\hrule}
height3pt&\omit&&\omit&&\omit&&\omit&&\omit&&\omit&&\omit&&\omit&&\omit&\cr
&\hfil 10\hfil&&\hfil 1.00\hfil&&\hfil 1.00\hfil&&\hfil 2.6\hfil&&\hfil 2.6\hfil&&
\hfil 5.3\hfil&&\hfil 5.3\hfil&&\hfil 26 \hfil&&\hfil 26\hfil&\cr
\noalign{\hrule}
height3pt&\omit&&\omit&&\omit&&\omit&&\omit&&\omit&&\omit&&\omit&&\omit&\cr
&\hfil 50\hfil&&\hfil 4.8\hfil&&\hfil 0.96\hfil&&\hfil 12.8\hfil&&\hfil 2.6\hfil&&
\hfil 26.4\hfil&&\hfil 5.3\hfil&&\hfil 148 \hfil&&\hfil 30\hfil&\cr
\noalign{\hrule}
height3pt&\omit&&\omit&&\omit&&\omit&&\omit&&\omit&&\omit&&\omit&&\omit&\cr
&\hfil 100\hfil&&\hfil 8.2\hfil&&\hfil 0.82\hfil&&\hfil 20.6\hfil&&\hfil 2.1\hfil&&
\hfil 47.8\hfil&&\hfil 4.8\hfil&&\hfil 234 \hfil&&\hfil 24\hfil&\cr
\noalign{\hrule}
height3pt&\omit&&\omit&&\omit&&\omit&&\omit&&\omit&&\omit&&\omit&&\omit&\cr
&\hfil 200\hfil&&\hfil 12.2\hfil&&\hfil 0.61\hfil&&\hfil 33.3\hfil&&\hfil 1.7\hfil&&
\hfil 65\hfil&&\hfil 3.3\hfil&&\hfil 282 \hfil&&\hfil 14\hfil&\cr
\noalign{\hrule}
height3pt&\omit&&\omit&&\omit&&\omit&&\omit&&\omit&&\omit&&\omit&&\omit&\cr
&\hfil 300\hfil&&\hfil 11.3\hfil&&\hfil 0.4\hfil&&\hfil 31\hfil&&\hfil 1.0\hfil&&
\hfil 64\hfil&&\hfil 2.1\hfil&&\hfil 262 \hfil&&\hfil 8.7\hfil&\cr
\noalign{\hrule}
height3pt&\omit&&\omit&&\omit&&\omit&&\omit&&\omit&&\omit&&\omit&&\omit&\cr
&\hfil 500\hfil&&\hfil 9.2\hfil&&\hfil 0.18\hfil&&\hfil 23\hfil&&\hfil 0.46\hfil&&
\hfil 45\hfil&&\hfil 0.9\hfil&&\hfil 195 \hfil&&\hfil 3.9\hfil&\cr
\noalign{\hrule}
height3pt&\omit&&\omit&&\omit&&\omit&&\omit&&\omit&&\omit&&\omit&&\omit&\cr
&\hfil 1000\hfil&&\hfil 3.1\hfil&&\hfil 0.031\hfil&&\hfil 7.7\hfil&&\hfil 0.08\hfil&&
\hfil 15.6\hfil&&\hfil 0.16\hfil&&\hfil 64 \hfil&&\hfil 0.64\hfil&\cr
\noalign{\hrule}
\noalign{\hrule}\noalign{\smallskip}
\multispan{19}Higgs-like scenario for the DM\hfil\hfil\cr
\multispan{19}$M=1000MeV=M_{exch};\beta_1 = 2\times 10^{-3};\theta_q = 0~(HOC)$\hfil\hfil\cr
\multispan{19}$E_\gamma^{min}=1\times 10^{-12}MeV;M_\gamma = 1\times 10^{-15}MeV;
RMAX=2\times 10^{-6}$\hfil\cr
\multispan{19}For 1000-200-1000: $M_\gamma = 10^{-15}MeV\to \sigma = 15027~ar.units~and$ \hfil\cr
\multispan{19}\phantom{For 1000-200-1000: }$M_\gamma = 10^{-14}MeV\to \sigma = 15074~ar.units$\hfil\cr
\multispan{19}$M_{_{DM}} = 1000.01~at~\beta_2 = \beta_1~in~order~to~avoid~a~resonance$\hfil\cr}}
$$}
\vfil\vfil\eject
\vfil\vfil\eject
\phantom{sssssss}
\vfil\vfil\eject
{\hsize=4.0truein
$$
\vbox{\offinterlineskip
\halign{&\vrule #&\strut\ #\ \cr
\multispan{19}\hfil\bf Table 2C\hfil\cr
\hfil\cr
\noalign{\smallskip}
\noalign{\hrule}
&\omit&&
\multispan{3}\hfil$\beta_2=1\times 10^{-3}$\hfil&&
\multispan{3}\hfil$\beta_2=2\times 10^{-3}$\hfil&&
\multispan{3}\hfil$\beta_2=3\times 10^{-3}$\hfil&&
\multispan{3}\hfil$\beta_2=6\times 10^{-3}$\hfil&\cr
\noalign{\hrule}
height3pt&\omit&&\omit&&\omit&&\omit&&\omit&&\omit&&\omit&&\omit&&\omit&\cr
&\hfil $M_{_{DM}}~(keV)$\hfil&&\hfil$R_\sigma$\hfil&&\hfil$R_{\rho\sigma}$\hfil&&
\hfil$R_\sigma$\hfil&&\hfil$R_{\rho\sigma}$\hfil&&\hfil$R_\sigma$\hfil&&\hfil $R_{\rho\sigma}$\hfil&&
\hfil$R_\sigma$\hfil&&\hfil$R_{\rho\sigma}$\hfil&\cr
height3pt&\omit&&\omit&&\omit&&\omit&&\omit&&\omit&&\omit&&\omit&&\omit&\cr
\noalign{\hrule}
height3pt&\omit&&\omit&&\omit&&\omit&&\omit&&\omit&&\omit&&\omit&&\omit&\cr
&\hfil 100\hfil&&\hfil 0.0018\hfil&&\hfil 0.018\hfil&&\hfil 0.0047\hfil&&\hfil 0.047\hfil&&
\hfil 0.010\hfil&&\hfil 0.10\hfil&&\hfil 0.048 \hfil&&\hfil 0.48\hfil&\cr
\noalign{\hrule}
height3pt&\omit&&\omit&&\omit&&\omit&&\omit&&\omit&&\omit&&\omit&&\omit&\cr
&\hfil 500\hfil&&\hfil 0.34\hfil&&\hfil 0.675\hfil&&\hfil 0.875\hfil&&\hfil 1.75\hfil&&
\hfil 1.91\hfil&&\hfil 3.8\hfil&&\hfil 8.66 \hfil&&\hfil 17.3\hfil&\cr
\noalign{\hrule}
height3pt&\omit&&\omit&&\omit&&\omit&&\omit&&\omit&&\omit&&\omit&&\omit&\cr
&\hfil 1000\hfil&&\hfil 1.00\hfil&&\hfil 1.00\hfil&&\hfil 2.74\hfil&&\hfil 2.74\hfil&&
\hfil 5.70\hfil&&\hfil 5.70\hfil&&\hfil 25.8 \hfil&&\hfil 25.8\hfil&\cr
\noalign{\hrule}
height3pt&\omit&&\omit&&\omit&&\omit&&\omit&&\omit&&\omit&&\omit&&\omit&\cr
&\hfil 2000\hfil&&\hfil 1.64\hfil&&\hfil 0.82\hfil&&\hfil 4.26\hfil&&\hfil 2.13\hfil&&
\hfil 8.8\hfil&&\hfil 4.4\hfil&&\hfil 39.4 \hfil&&\hfil 19.7\hfil&\cr
\noalign{\hrule}
height3pt&\omit&&\omit&&\omit&&\omit&&\omit&&\omit&&\omit&&\omit&&\omit&\cr
&\hfil 3000\hfil&&\hfil 1.69\hfil&&\hfil 0.56\hfil&&\hfil 4.38\hfil&&\hfil 1.46\hfil&&
\hfil 9.0\hfil&&\hfil 3.0\hfil&&\hfil 40.8 \hfil&&\hfil 13.6\hfil&\cr
\noalign{\hrule}
height3pt&\omit&&\omit&&\omit&&\omit&&\omit&&\omit&&\omit&&\omit&&\omit&\cr
&\hfil 5000\hfil&&\hfil 1.47\hfil&&\hfil 0.29\hfil&&\hfil 3.77\hfil&&\hfil 0.75\hfil&&
\hfil 7.8\hfil&&\hfil 1.56\hfil&&\hfil 35.3 \hfil&&\hfil 7.1\hfil&\cr
\noalign{\hrule}
\noalign{\hrule}\noalign{\smallskip}
\multispan{19}DM~+~Exchange particle scenario\hfil\hfil\cr
\multispan{19}$M=511keV;M_{exch}=100keV;\beta_1 = 2\times 10^{-3};\theta_q = 0~(HOC)$\hfil\hfil\cr
\multispan{19}$E_\gamma^{min}=1\times 10^{-9}keV;M_\gamma = 1\times 10^{-16}keV;
RMAX=2\times 10^{-4}$\hfil\cr
\multispan{19}For 511-1000-100: $M_\gamma = 10^{-16}keV\to\sigma = 53477869~ar.units~and$\hfil\cr
\multispan{19}\phantom{For 511-1000-100: }$M_\gamma = 10^{-15}keV\to\sigma = 53477861~ar.units$\hfil\cr}}
$$}
%$$}
{\hsize=4.0truein
$$
\vbox{\offinterlineskip
\halign{&\vrule#&\strut\ #\ \cr
\multispan{19}\hfil\bf Table 2D\hfil\cr
\hfil\cr
\noalign{\smallskip}
\noalign{\hrule}
&\omit&&
\multispan{3}\hfil$\beta_2=1\times 10^{-3}$\hfil&&
\multispan{3}\hfil$\beta_2=2\times 10^{-3}$\hfil&&
\multispan{3}\hfil$\beta_2=3\times 10^{-3}$\hfil&&
\multispan{3}\hfil$\beta_2=6\times 10^{-3}$\hfil&\cr
\noalign{\hrule}
height3pt&\omit&&\omit&&\omit&&\omit&&\omit&&\omit&&\omit&&\omit&&\omit&\cr
&\hfil$M_{_{DM}}~(MeV)$\hfil&&\hfil$R_\sigma$\hfil&&\hfil$R_{\rho\sigma}$\hfil&&
\hfil$R_\sigma$\hfil&&\hfil$R_{\rho\sigma}$\hfil&&\hfil$R_\sigma$\hfil&&\hfil $R_{\rho\sigma}$\hfil&&
\hfil$R_\sigma$\hfil&&\hfil$R_{\rho\sigma}$\hfil&\cr
height3pt&\omit&&\omit&&\omit&&\omit&&\omit&&\omit&&\omit&&\omit&&\omit&\cr
\noalign{\hrule}
height3pt&\omit&&\omit&&\omit&&\omit&&\omit&&\omit&&\omit&&\omit&&\omit&\cr
&\hfil 500\hfil&&\hfil 0.04\hfil&&\hfil 0.25\hfil&&\hfil 0.11\hfil&&\hfil 0.64\hfil&&
\hfil 0.22\hfil&&\hfil 1.35\hfil&&\hfil 1.0 \hfil&&\hfil 6.0\hfil&\cr
\noalign{\hrule}
height3pt&\omit&&\omit&&\omit&&\omit&&\omit&&\omit&&\omit&&\omit&&\omit&\cr
&\hfil 1000\hfil&&\hfil 0.25\hfil&&\hfil 0.75\hfil&&\hfil 0.63\hfil&&\hfil 1.89\hfil&&
\hfil 1.31\hfil&&\hfil 3.9\hfil&&\hfil 5.9 \hfil&&\hfil 17.6\hfil&\cr
\noalign{\hrule}
height3pt&\omit&&\omit&&\omit&&\omit&&\omit&&\omit&&\omit&&\omit&&\omit&\cr
&\hfil 2000\hfil&&\hfil 0.73\hfil&&\hfil 1.1\hfil&&\hfil 1.88\hfil&&\hfil 2.8\hfil&&
\hfil 3.84\hfil&&\hfil 5.75\hfil&&\hfil 16.4 \hfil&&\hfil 24.6\hfil&\cr
\noalign{\hrule}
height3pt&\omit&&\omit&&\omit&&\omit&&\omit&&\omit&&\omit&&\omit&&\omit&\cr
&\hfil 3000\hfil&&\hfil 1.00\hfil&&\hfil 1.00\hfil&&\hfil 2.54\hfil&&\hfil 2.54\hfil&&
\hfil 5.2\hfil&&\hfil 5.2\hfil&&\hfil 22 \hfil&&\hfil 22\hfil&\cr
\noalign{\hrule}
height3pt&\omit&&\omit&&\omit&&\omit&&\omit&&\omit&&\omit&&\omit&&\omit&\cr
&\hfil 5000\hfil&&\hfil 1.14\hfil&&\hfil 0.68\hfil&&\hfil 2.9\hfil&&\hfil 1.7\hfil&&
\hfil 6.0\hfil&&\hfil 3.6\hfil&&\hfil 25 \hfil&&\hfil 15\hfil&\cr
\noalign{\hrule}
height3pt&\omit&&\omit&&\omit&&\omit&&\omit&&\omit&&\omit&&\omit&&\omit&\cr
&\hfil 10000\hfil&&\hfil 1.0\hfil&&\hfil 0.30\hfil&&\hfil 2.5\hfil&&\hfil 0.75\hfil&&
\hfil 5.0\hfil&&\hfil 1.50\hfil&&\hfil 20.8 \hfil&&\hfil 6.2\hfil&\cr
\noalign{\hrule}
\noalign{\hrule}\noalign{\smallskip}
\multispan{19}DM~+~Exchange particle scenario\hfil\hfil\cr
\multispan{19}$M=1000MeV;M_{exch}=100MeV;\beta_1 = 2\times 10^{-3};\theta_q = 0~(HOC)$\hfil\hfil\cr
\multispan{19}$E_\gamma^{min}=1\times 10^{-10}MeV;M_\gamma = 1\times 10^{-15}MeV;
RMAX=1\times 10^{-4}$\hfil\cr
\multispan{19}For 1000-3000-100: $M_\gamma = 10^{-15}MeV\to \sigma = 390753 ~ar.units~and$ \hfil\cr
\multispan{19}\phantom{For 1000-3000-100: }$M_\gamma = 10^{-14}MeV\to \sigma = 390749~ar.units$\hfil\cr
\multispan{19}$M_{_{DM}} = 1000.01~at~\beta_2 = \beta_1~in~order~to~avoid~a~resonance$\hfil\cr}}
$$}
\vfil\vfil\eject
\phantom{sssss}
\vfil\vfil\eject
\noindent
{\hsize=4.0truein
$$
\vbox{\offinterlineskip
\halign{&\vrule#&\strut\ #\ \cr
\multispan{13}\hfil\bf Table 3A\hfil\cr
\multispan{13}\hfil $number~of~photons~/keV/m^2/sec~at~21cm~(10^{-6}/keV/m^2/sec)$\hfil\cr
\multispan{13}\hfil in the DM-exchange particle scenario
\hfil\cr
\noalign{\smallskip}
\noalign{\hrule}
height3pt&\omit&&\omit&&\omit&&\omit&&\omit&&\omit&\cr
&\hfil matter-DM-exchange \hfil&&\hfil$\theta=175$\hfil&&\hfil$\theta=135$\hfil&&
\hfil$\theta=90$\hfil&&\hfil$\theta=45$\hfil&&\hfil$\theta=5$\hfil&\cr
height3pt&\omit&&\omit&&\omit&&\omit&&\omit&&\omit&\cr
\noalign{\hrule}
height3pt&\omit&&\omit&&\omit&&\omit&&\omit&&\omit&\cr
&\hfil 511-1000-100(keV)\hfil&&\hfil 52\hfil&&\hfil 4.8\hfil&&
\hfil 1.8\hfil&&\hfil 1.12\hfil&&\hfil 0.96\hfil&\cr
height3pt&\omit&&\omit&&\omit&&\omit&&\omit&&\omit&\cr
\noalign{\hrule}
height3pt&\omit&&\omit&&\omit&&\omit&&\omit&&\omit&\cr
&\hfil 511-3000-100(keV)\hfil&&\hfil 13\hfil&&\hfil 1.2\hfil&&
\hfil 0.63\hfil&&\hfil 0.28\hfil&&\hfil 0.24\hfil&\cr
height3pt&\omit&&\omit&&\omit&&\omit&&\omit&&\omit&\cr
\noalign{\hrule}
height3pt&\omit&&\omit&&\omit&&\omit&&\omit&&\omit&\cr
&\hfil 511-100-100(keV)\hfil&&\hfil 1.7\hfil&&\hfil 0.17\hfil&&
\hfil 0.08\hfil&&\hfil 0.04\hfil&&\hfil 0.03\hfil&\cr
height3pt&\omit&&\omit&&\omit&&\omit&&\omit&&\omit&\cr
\noalign{\hrule}
height3pt&\omit&&\omit&&\omit&&\omit&&\omit&&\omit&\cr
&\hfil 1000-3000-100(MeV)\hfil&&\hfil 0.77\hfil&&\hfil 0.09\hfil&&
\hfil 0.04\hfil&&\hfil 0.025\hfil&&\hfil 0.02\hfil&\cr
height3pt&\omit&&\omit&&\omit&&\omit&&\omit&&\omit&\cr
\noalign{\hrule}
\noalign{\hrule}\noalign{\smallskip}}}
$$}
{\hsize=4.0truein
$$
\vbox{\offinterlineskip
\halign{&\vrule#&\strut\ #\ \cr
\multispan{13}\hfil\bf Table 3B\hfil\cr
\multispan{13}\hfil $number~of~photons /keV/m^2/sec~at~21 cm$ \hfil\cr
\multispan{13}\hfil in the Higgs-like scenario
\hfil\cr
\noalign{\smallskip}
\noalign{\hrule}
height3pt&\omit&&\omit&&\omit&&\omit&&\omit&&\omit&\cr
&\hfil matter-DM-exchange \hfil&&\hfil$\theta=175$\hfil&&\hfil$\theta=135$\hfil&&
\hfil$\theta=90$\hfil&&\hfil$\theta=45$\hfil&&\hfil$\theta=5$\hfil&\cr
height3pt&\omit&&\omit&&\omit&&\omit&&\omit&&\omit&\cr
\noalign{\hrule}
height3pt&\omit&&\omit&&\omit&&\omit&&\omit&&\omit&\cr
&\hfil 511-100-511(keV)\hfil&&\hfil $2.8\times 10^{-6}$\hfil&&\hfil $0.29\times 10^{-6}$\hfil&&
\hfil $0.13\times 10^{-6}$\hfil&&\hfil$0.06\times 10^{-6}$\hfil&&\hfil$0.056\times 10^{-6}$\hfil&\cr
height3pt&\omit&&\omit&&\omit&&\omit&&\omit&&\omit&\cr
\noalign{\hrule}
height3pt&\omit&&\omit&&\omit&&\omit&&\omit&&\omit&\cr
&\hfil 1000-200-1000(MeV)\hfil&&\hfil$100\times 10^{-17}$\hfil&&\hfil$9.2\times 10^{-17}$\hfil&&
\hfil$4.0\times 10^{-17}$\hfil&&\hfil$2.6\times 10^{-17}$\hfil&&\hfil$2.0\times 10^{-17}$\hfil&\cr
height3pt&\omit&&\omit&&\omit&&\omit&&\omit&&\omit&\cr
\noalign{\hrule}
height3pt&\omit&&\omit&&\omit&&\omit&&\omit&&\omit&\cr
&\hfil 1000-17-1000(MeV)\hfil&&\hfil$24\times 10^{-17}$\hfil&&\hfil$2.4\times 10^{-17}$\hfil&&
\hfil$1.2\times 10^{-17}$\hfil&&\hfil$0.72\times 10^{-17}$\hfil&&\hfil$0.48\times 10^{-17}$\hfil&\cr
height3pt&\omit&&\omit&&\omit&&\omit&&\omit&&\omit&\cr
\noalign{\hrule}
\noalign{\hrule}\noalign{\smallskip}}}
$$}
\vfil\vfil\eject
\phantom {ssss}
\vfil\vfil\eject
{\hsize=7.0truein
$$
\vbox{\offinterlineskip
\halign{&\vrule#&\strut\ #\ \cr
\multispan{13}\hfil\bf Table 4\hfil\cr
\multispan{13}\hfil $\lambda/(m_{exchange}/100{k\atop M}eV)~(\#\times 10^5)$\hfil\cr
\multispan{13}\hfil in the DM-exchange particle scenario
\hfil\cr
\noalign{\smallskip}
\noalign{\hrule}
height3pt&\omit&&\omit&&\omit&&\omit&&\omit&&\omit&\cr
&\hfil matter-DM-exchange \hfil&&\hfil$\theta=175$\hfil&&\hfil$\theta=135$\hfil&&
\hfil$\theta=90$\hfil&&\hfil$\theta=45$\hfil&&\hfil$\theta=5$\hfil&\cr
height3pt&\omit&&\omit&&\omit&&\omit&&\omit&&\omit&\cr
\noalign{\hrule}
height3pt&\omit&&\omit&&\omit&&\omit&&\omit&&\omit&\cr
&\hfil 511-1000-100(keV)\hfil&&\hfil 5\hfil&&\hfil 9\hfil&&
\hfil 12\hfil&&\hfil 13\hfil&&\hfil 14\hfil&\cr
height3pt&\omit&&\omit&&\omit&&\omit&&\omit&&\omit&\cr
\noalign{\hrule}
height3pt&\omit&&\omit&&\omit&&\omit&&\omit&&\omit&\cr
&\hfil 511-3000-100(keV)\hfil&&\hfil 7\hfil&&\hfil 13\hfil&&
\hfil 15\hfil&&\hfil 18\hfil&&\hfil 19\hfil&\cr
height3pt&\omit&&\omit&&\omit&&\omit&&\omit&&\omit&\cr
\noalign{\hrule}
height3pt&\omit&&\omit&&\omit&&\omit&&\omit&&\omit&\cr
&\hfil 511-100-100(keV)\hfil&&\hfil 12\hfil&&\hfil 21\hfil&&
\hfil 25\hfil&&\hfil 30\hfil&&\hfil 32\hfil&\cr
height3pt&\omit&&\omit&&\omit&&\omit&&\omit&&\omit&\cr
\noalign{\hrule}
height3pt&\omit&&\omit&&\omit&&\omit&&\omit&&\omit&\cr
&\hfil 1000-3000-100(MeV)\hfil&&\hfil 14\hfil&&\hfil 24\hfil&&
\hfil 29\hfil&&\hfil 33\hfil&&\hfil 35\hfil&\cr
height3pt&\omit&&\omit&&\omit&&\omit&&\omit&&\omit&\cr
\noalign{\hrule}
\noalign{\hrule}\noalign{\smallskip}
\multispan{13}Constraints to be expected from the detectors\hfil\hfil\hfil\cr}}
$$}
\vfil\vfil\eject
\phantom{ssssss}
\end{document}